\documentclass{article}
\usepackage{spconf}
\usepackage{graphicx}
\usepackage{stfloats}
\usepackage{amsmath,amssymb}
\usepackage{mathtools}
\usepackage{cite}
\usepackage{url}
\usepackage{xcolor}
\usepackage{soul}
\usepackage{comment}
\usepackage[caption=false,font=footnotesize]{subfig}
\usepackage{balance}
\usepackage[pscoord]{eso-pic}

\def\E{\mathbb{E}}
\def\x{\boldsymbol{x}}
\def\y{\boldsymbol{y}}
\def\q{\boldsymbol{q}}
\def\v{\boldsymbol{v}}
\def\V{\boldsymbol{V}}
\def\w{\boldsymbol{w}}
\def\A{\boldsymbol{A}}
\def\F{{\mathcal{F}}}

\def\F{\mathcal{F}}
\def\M{\mathcal{M}}
\def\Fviral{\F \text{ viral}}
\newcommand{\normaldensity}[3]{\mathcal{N}({#1}; {#2}, {#3})}

\def\pif{\pi_{\text{vf}}}

\def\pih{\pi_{\text{vf}}}
\def\pip{\pi_{\text{ind}}}

\def\lasso{\textsc{Lasso}}
\def\sqrtglasso{\textsc{Sqrt-Glasso}}
\def\sqrtoglasso{\textsc{Sqrt-Oglasso}}
\def\comp{\textsc{Comp}}
\def\complasso{\textsc{Comp-Lasso}}
\def\compsqrtglasso{\textsc{Comp-Sqrt-Glasso}}
\def\compsqrtoglasso{\textsc{Comp-Sqrt-Oglasso}}

\graphicspath{{./figures_main/}{./figures_sup/}}

\newcommand{\placetextbox}[3]{% \placetextbox{<horizontal pos>}{<vertical pos>}{<stuff>}
  \setbox0=\hbox{#3}% Put <stuff> in a box
  \AddToShipoutPictureFG*{% Add <stuff> to current page foreground
    \put(\LenToUnit{#1\paperwidth},\LenToUnit{#2\paperheight}){\vtop{{\null}\makebox[0pt][c]{#3}}}%
  }%
}%

\title{Contact Tracing Enhances the Efficiency of COVID-19 Group Testing}

\name{Ritesh Goenka,$^{\star1}$ 
Shu-Jie Cao,$^{\star2}$ 
Chau-Wai Wong,$^3$ 
Ajit Rajwade,$^1$ 
Dror Baron$^3$
\thanks{$^{\star}$\,RG and SJC have made equal contributions.
AR acknowledges support from SERB Grant \#10013890, IITB-WRCB Grant \#DONWR04-002, and DST-Rakshak grant \#DST0000-005. 
CWW acknowledges support from NSF Grant \#2030430.
The supplemental document can be found on the webpages of the last three authors.
\vspace{-0mm}}}
\address{$^1$\,IIT Bombay, India,\hspace{3mm}
$^2$\,ShanghaiTech University, China,\hspace{3mm} 
$^3$\,North Carolina State University, USA}

\begin{document}

\maketitle

\begin{abstract}
Group testing can save testing resources in the context of the ongoing COVID-19 pandemic. In group testing, we are given $n$ samples, one per individual, and arrange them into $m < n$ pooled samples, where each pool is obtained by mixing a subset of the $n$ individual samples. Infected individuals are then identified using a group testing algorithm. In this paper, we use side information (SI) collected from contact tracing (CT) within non-adaptive/single-stage group testing algorithms. We generate data by incorporating CT SI and characteristics of disease spread between individuals. These data are fed into two signal and measurement models for group testing, where numerical results show that our algorithms provide improved sensitivity and specificity. While Nikolopoulos et al. utilized family structure to improve non-adaptive group testing, ours is the first work to explore and demonstrate how CT SI can further improve group testing performance.
\end{abstract}


%=====

\vspace{-2mm}
\section{Introduction}
\label{sec:intro}
\vspace{-1mm}
Widespread testing has been promoted for combating the ongoing COVID-19 pandemic.
Samples are typically collected from nasal or oropharyngeal swabs, and then processed by a reverse transcription polymerase chain reaction (RT-PCR) machine.
However, widespread testing is hindered by supply chain constraints and long testing times. 

Pooled or {\em group testing} has been suggested for improving testing efficiencies \cite{Abdelhamid2020}.
Group testing involves mixing a subset of $n$ individual samples into $m < n$ pools. The measurement process can be expressed as $\boldsymbol{y} = \mathfrak{N}(\boldsymbol{Ax})$,
where $\boldsymbol{x}$ is a vector that quantifies the health status of the $n$ individuals,
$\boldsymbol{A}$ is an $m \times n$ binary pooling matrix with $A_{ij} = 1$ if the $j$th individual contributes to the $i$th pool, else $A_{ij} = 0$,
$\boldsymbol{y}$ is a vector of $m$ noisy measurements or tests,
and $\mathfrak{N}$ represents a probabilistic 
noise model that relates the noiseless pooled results, $\boldsymbol{Ax}$, to 
$\boldsymbol{y}$.
We consider two signal and noise models.\\
{\bf Model \textbf{M1}:} A {\em binary noise} model used by Zhu et al.~\cite{Zhu2020}, 
where $\boldsymbol{x}$ is binary,
$\boldsymbol{w} = \boldsymbol{Ax}$
is an auxiliary vector,
and the measurement $y_i\in\{0,1\}$ depends probabilistically on $w_i$,
where $\Pr(y_i=1|w_i=0)$ and $\Pr(y_i=0|w_i>0)$ are probabilities of erroneous tests.

\noindent{}{\bf Model \textbf{M2}:} A {\em multiplicative noise} model of the form 
$\boldsymbol{y} = \boldsymbol{Ax} \circ \boldsymbol{z}$ as used in Ghosh et al.~\cite{Ghosh2020},
where $\circ$ represents element-wise multiplication, $\boldsymbol{z}$ is a vector of $m$ noisy elements defined as
$z_i = (1+q)^{\eta_i}$, $q \in (0, 1]$ is a known amplification factor for RT-PCR, 
$\eta_i \sim \mathcal{N}(0,\sigma^2)$, and $\sigma^2 \ll 1$ is a known parameter controlling the strength of the noise in RT-PCR. Under model \textbf{M2}, $\boldsymbol{x}$ and $\boldsymbol{y}$ represent viral loads in the 
$n$ individuals and $m$ pools, respectively. Assuming reasonably high viral loads in $\boldsymbol{x}$, 
Poisson effects in $\boldsymbol{y}$ can be ignored \cite{Ghosh2020}.

For both models, the objective is to estimate $\boldsymbol{x}$ from $\boldsymbol{y}$ and $\boldsymbol{A}$.
We use single-stage {\em non-adaptive} algorithms as in \cite{Zhu2020, Ghosh2020}, rather than two-stage algorithms,
which employ a second stage of tests depending on results from the first stage, as in 
Heidarzadeh and Narayanan~\cite{Heiderzadeh2020} or the classical approach by Dorfman~\cite{Dorfman1943}. 
The main advantage of non-adaptive algorithms is that they save on testing time, which is quite high for RT-PCR. 
Algorithms for estimation of $\boldsymbol{x}$ from $\boldsymbol{y}$ and $\boldsymbol{A}$~\cite{Ghosh2020, Shental2020} 
rely primarily on the {\em sparsity} of $\boldsymbol{x}$, which is a valid assumption for COVID-19 due to low prevalence rates \cite{Benatia2020}.
Zhu et al.~\cite{Zhu2020} also exploit {probabilistic} information such as the prevalence rate, structure in $\boldsymbol{x}$, and {\em side information} (SI).
Finally, Nikolopoulos et al.~\cite{Nikolopoulos2020} showed how SI about 
{family-style structure} can reduce the number of required tests, $m$.

In this paper, we show how to estimate $\boldsymbol{x}$ while utilizing {\em contact tracing} (CT) SI, which allows one to analyze the spread of the pandemic \cite{cdc_contact_tracing}.
Our contributions are twofold.
First, we propose a generative model for a population of $n$ individuals that characterizes the spread of COVID-19 by explicitly using CT SI.
Second, we show that CT SI, when used appropriately, can help algorithms such as generalized approximate message passing~(GAMP)~\cite{rangan2011generalized}
or LASSO variants \cite{Yuan2006,Jacot2020}
better estimate $\boldsymbol{x}$ from $\boldsymbol{y}$ and $\boldsymbol{A}$.
Our work uses more SI than Nikolopoulos et al.~\cite{Nikolopoulos2020}, who only considered family-style structure in binary group testing.

%=====

\vspace{-2mm}
\section{Data Generation Model}
\label{sec:data_gen}
\vspace{-1mm}
In this section, we present a generative infection model incorporating CT SI, 
which we later use to prepare simulated data for algorithmic evaluation.
We model a population of $n$ individuals using a dynamical or time-varying graphical model that contains nodes $\{v_i\}_{i=1}^n$ and undirected edges $\big\{e_{ij}^{(t)}\big\}_{i,j=1}^n$.
On a given day $t$, an edge $e_{ij}^{(t)}$ between nodes $v_i$ and $v_j$ encodes CT SI 
$\big(\tau^{(t)}_{ij}, d^{(t)}_{ij}\big)$, which can be acquired via Bluetooth-based CT applications~\cite{Hekmati2020}. 
Here, $\tau^{(t)}_{ij}$ represents the contact duration and $d^{(t)}_{ij}$ represents a measure of the physical proximity
between two individuals.
On day~$t$, a node can be in one of the following states: \emph{susceptible}, \emph{infected}, \emph{infectious}, and \emph{recovered}.
To keep the model simple, we assume that there are no reinfections, i.e., recovered is a terminal state, despite some reports of reinfection\cite{Haseltine2020}.
\begin{figure}[!t]
  \vspace{-1mm}
  \centering
  \centerline{\includegraphics[width=8cm]{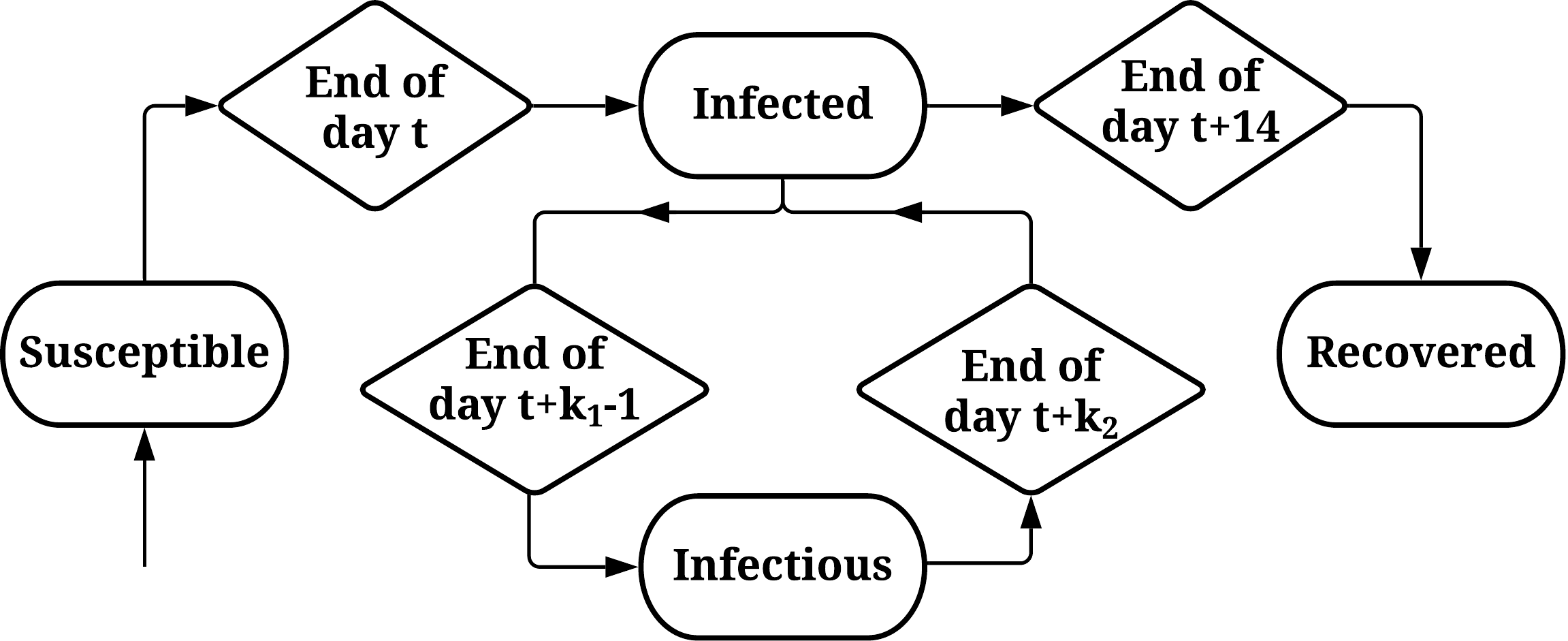}}
  \vspace{-4mm}
  \caption{State transition diagram for a node. A node is infectious only between days $k_1$ and $k_2$ (both inclusive) 
after getting infected. In our work, we set $(k_1,k_2) = (3,7)$.}
\label{fig:state-transition-diagram}
\vspace{-4mm}
\end{figure}

We adopt a simplified infection dynamic wherein the
infectious period is preceded and followed by the infected state.
We propose the following design parameters for the infection dynamics based on a World Health Organization report on COVID-19~\cite{WHOreport}.
Specifically, a node $v_i$ remains infected but noninfectious for $k_1 = 3$ days.
On day $t+k_1$, the node becomes infectious and may transmit the disease to a susceptible neighboring node $v_j$ 
with probability $p_{i,j}^{(t+k_1)}$ whose construction is described below.
An infectious node can potentially transmit the infection until $k_2 = 7$ days after getting infected, and becomes noninfectious afterward.
We also model a small fraction of stray infections that may occur, for example, due to sporadic contact with contaminated surfaces. 
Such infections only affect nodes in the susceptible state with a probability $p_1 = 2 \times 10^{-4}$
of our choice.
A state diagram for a general node is illustrated in Fig.~\ref{fig:state-transition-diagram}. 
Regarding the viral load $x_i^{(t)}$ for node $i$ on day $t$, we assume $x_i^{(t)} = 0$ if the node is susceptible or recovered.
For an infected or infectious node, we make a simplified assumption that its viral load $x_i^{(t)} \sim \textrm{Uniform}(1,32768)$,\footnote{
The cycle threshold for RT-PCR commonly ranges from $19$ to $34$ cycles \cite[Fig. 3]{Buchan2020}, where $34$ cycles corresponds to a low initial viral load of a few molecules, and each cycle roughly doubles the viral density.
Therefore, we estimate the largest possible viral load as $2^{34-19} = 2^{15} = 32768$.}
once drawn, remains constant throughout the combined $14$-day period of infection.

Next, we model the probability $p_{i,j}^{(t)}$ that the disease is transmitted from node $v_i$ to $v_j$ on day $t$.
We view the infection times of the population throughout the pandemic as 
a nonhomogeneous Poisson process with time-varying rate function $\lambda(t)$.
Consider a $\tau^{(t)}_{ij}$-hour contact on day $t$ when susceptible node $v_j$ is exposed to infectious node $v_i$. 
The average infection rate $\lambda_{ij}(t)$ for day $t$ is assumed to be proportional to both the viral load $x^{(t)}_i$ and the physical proximity $d^{(t)}_{ij}$, namely, $\lambda_{ij}(t) = \lambda_0 \, x^{(t)}_i \, d^{(t)}_{ij}$, where $\lambda_0$ is a tunable, baseline Poisson rate.
The probability that $v_j$ is infected by the end of contact period $\tau^{(t)}_{ij}$ is therefore $p^{(t)}_{i,j} = 1 - \exp \left( -\lambda_0 \, x^{(t)}_i \, d^{(t)}_{ij} \, \tau^{(t)}_{ij} \right)$ for $t \in [k_1, k_2] + t_i$.
From the standpoint of susceptible node $v_j$, all its neighbors $v_k$ that are infectious contribute to its probability of getting infected on day~$t$,
namely,  
$1-\prod_{k} \big( 1 - p^{(t)}_{k,j} \big)$.

\begin{figure}[!t]
  \vspace{-5mm}
  \subfloat[width=0.48\linewidth][]{\includegraphics[width=4.3cm]{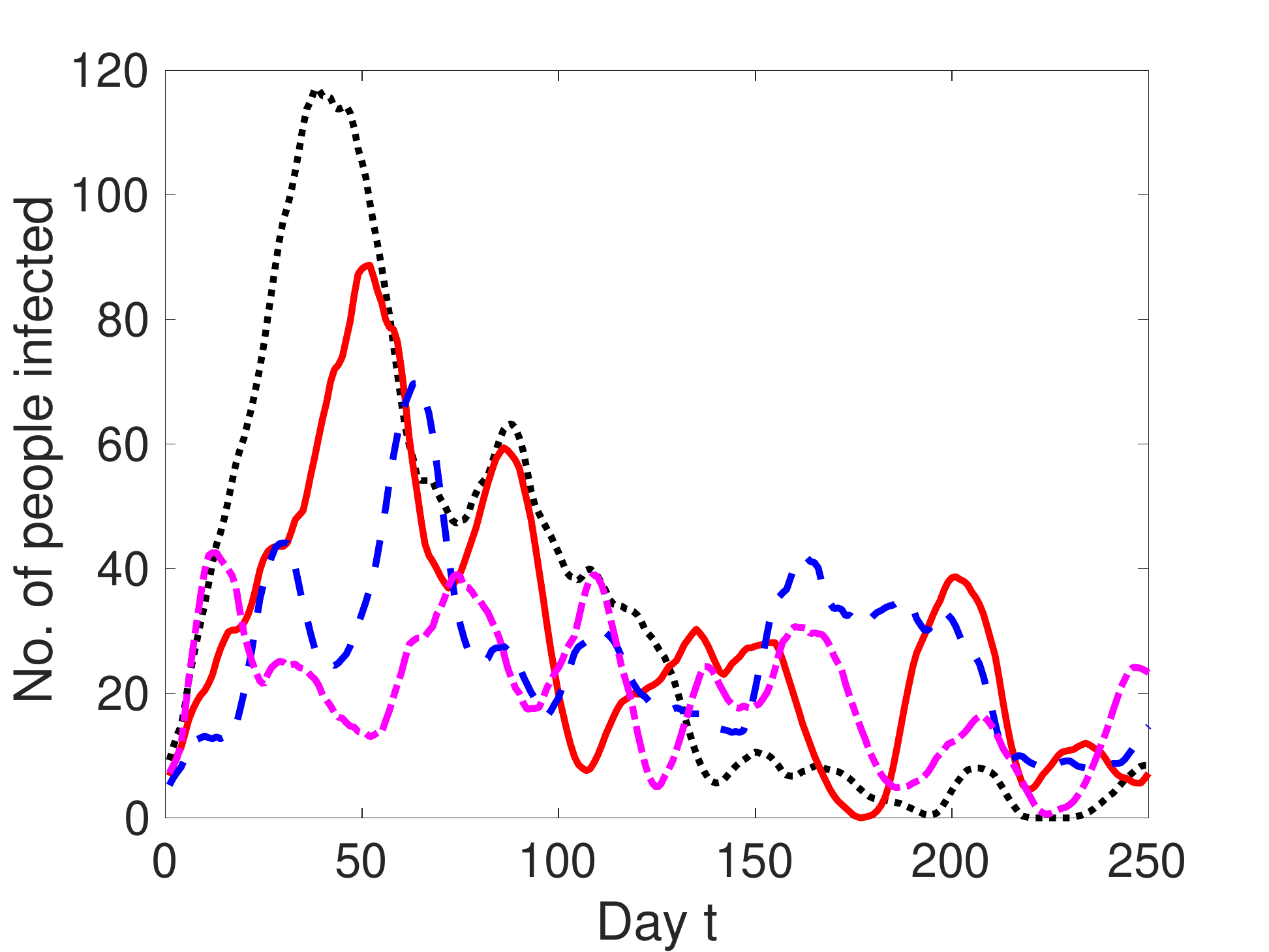}}
  \subfloat[width=0.48\linewidth][]{\includegraphics[width=4.3cm]{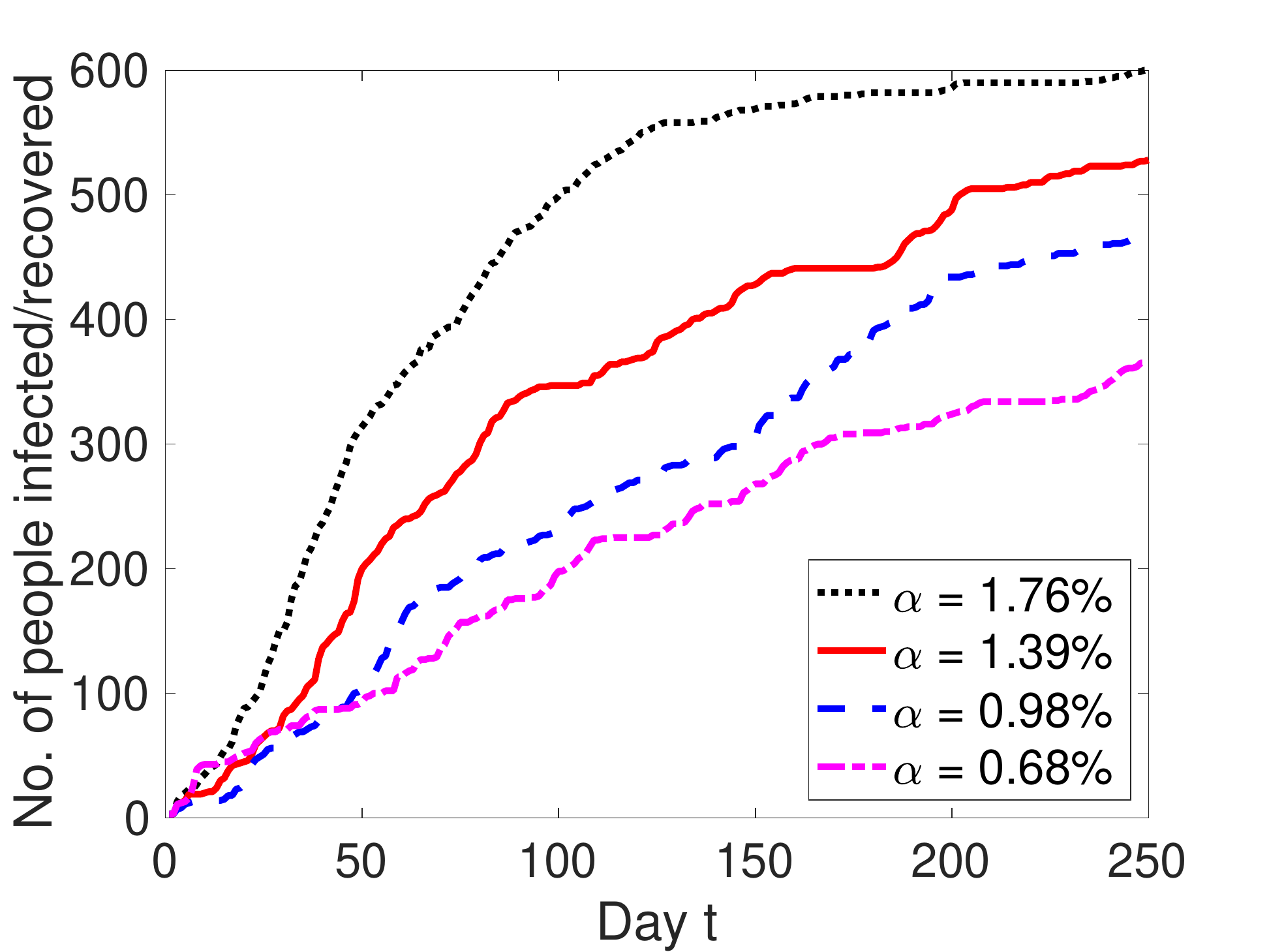}}
  \vspace{-3mm}
  \caption{(a) The number of active infections, and (b) cumulative infections at different inter-clique contact levels $\alpha$. We chose $50$-day windows for testing proposed algorithms.}
\label{fig:infections-curves}
\vspace{-5mm}
\end{figure}

While generating our data, we considered $n = 1000$ nodes divided into cliques based on the distribution of family sizes in India \cite[pg. 18]{UNDoc}, for a duration of $t_{\text{max}} = 250$ days.
Fig.~\ref{fig:infections-curves} shows the number of active infections and the cumulative number of infections at the end of each day.
The clique structures were kept constant throughout the $t_{\text{max}}$ days, whereas inter-clique contacts corresponding to sporadic contacts between people were dynamically added and removed. 
The varying inter-clique contact level $\alpha$ affects the sparsity of the underlying vector $\boldsymbol{x}$ as it brings infections to new cliques/families.
Pooling of samples is performed at the beginning of each day from $t_{\text{peak}}-24$ to $t_{\text{peak}}+25$, 
where $t_{\text{peak}}$ is the day with the maximum number of active infections.

%=====

\vspace{-2mm}
\section{Proposed Group Testing Algorithms}
\label{sec:algos}
\vspace{-1mm}
This section describes two classes of group testing algorithms for reconstructing the health status vector $\x$ from the pooled tests, $\y$, and pooling matrix, $\A$.

\noindent{}{\bf Algorithms for binary noise.}
For model \textbf{M1}, Zhu et al.~\cite{Zhu2020} use
{\em generalized approximate message passing} (GAMP)~\cite{rangan2011generalized} 
for group testing estimation.
GAMP is comprised of two components.
The first component consists of
an input channel that relates a prior for $n$ individuals' viral loads, 
$\x=(x_i)_{i=1}^n$, and pseudo data, $\v = \x + \q \in \mathbb{R}^n$, 
where
the $n$ coordinates of $\x$ are correlated, and
$\q$ is additive white Gaussian noise with $q_i \sim \mathcal{N}(0,\Delta)$.
We estimate $\x$ from $\v$ using a denoising function, {often called a denoiser,}
\begin{equation}
\widehat{x}_{i}=g_{\text{in}} \left(\v \right) 
= \E \left[X_{i} \mid \V=\v \right] \label{input_d},
\vspace*{-2mm}
\end{equation}
where we use the convention that when both the upper and lower case versions
of a symbol appear, the upper case is a random variable and the lower case its realization, and
$\E \left[X_{i} | \v \right]$ represents $\E \left[X_{i} | \V=\v \right]$ when the context is clear.
The second component of GAMP consists of an output channel relating the auxiliary vector $\w$ to the noisy measurements $\y$ as reviewed in Sec.~\ref{sec:intro}.
We adopt the output channel {denoiser} of Zhu et al.~\cite{Zhu2020}, 
$h_{i}=g_{\text{out}}\left(y_{i}; k_{i}, \theta_{i}\right) = ( \E\left[W_{i} \mid y_{i}, k_{i},  \theta_{i}\right]-k_{i} ) / \theta_i$, 
where $\theta_{i}$ is the estimated variance of $h_i$, 
and $k_{i}$ is the mean of our estimate for $w_i$. 
Since $y_i$ depends probabilistically on $w_i$, we have 
$f \left(w_{i} \mid y_{i}, k_{i}, \theta_{i}\right) \propto \operatorname{Pr}\left(y_{i} \mid w_{i}\right) \, 
\exp \left[-\frac{\left(w_{i}-k_{i}\right)^{2}}{2 \theta_{i}}\right]$,
where $W_i$ is approximated as Gaussian in the derivation of GAMP.

While Zhu et al.~\cite{Zhu2020} considered Bernoulli $\x$,
which implies a scalar separable {denoiser} $g_{\text{in}}$ for the input channel, 
this paper accounts for probabilistic dependencies within $\x$.
Our first probabilistic model considers groups of people, for example, members of a family.
Each family
is modeled as entirely healthy with probability $1-\pih$,
else each individual within the family is infected with probability $\pip$.
This model relates to our generative model of Sec.~\ref{sec:data_gen} by using
family structure as SI.
Denoting the pseudo data of family~$\F$ by $\v_{\F}$,
the denoiser for the $i$th individual of
family~$\F$ is given by
\vspace*{-2mm}
\begin{equation}
\small
g_{\text{in}}^\text{family}(\v_{\F} ) 
= \E\big[ X_i | \Fviral,\v_{\F} \big] \, \Pr\big(\Fviral|\v_{\F} \big),
\normalsize
\label{eq:denoiser_f}
\vspace*{-1mm}
\end{equation}
where $\E\big[X_i|\Fviral,\v_{\F} \big]$ and $\Pr \big( \Fviral|\v_{\F} \big)$
are parameterized by $\pih$, $\pip$, and $\Delta$.
For detailed expressions, we refer readers to Sec.~1.1 of the supplemental document.

Our second probabilistic model uses CT.
Consider a hypothetical widespread testing program that relies on CT SI,
where all individuals are tested $8$ days before the group testing program begins
resulting in a good estimate of their ground-truth health status.
After the program begins, probability estimates from the previous group test are used as priors for the $n$ individuals when performing the current group test.
We provide detailed analysis in Secs.~2.2--2.3 of the supplemental document on the use of prior infection status.
The final form of the denoiser for the CT model is as follows:
\vspace*{-2mm}
\begin{equation}
\small
g_{\text{in}}^\text{CT}(v_i)
\!=\! \left\{ 1 \!+\! \big[\Pr(X_i\!=\!1)^{-1} \! - \! 1\big] \exp \Big[ \big(v_i\!-\!\tfrac{1}{2}\big) \big/ \Delta \Big] \!\right\}^{-1}\!.
\label{eq:denoiser_ct}
\vspace*{-1mm}
\end{equation}
\normalsize
Here, $\Pr(X_i\!=\!1)$ for day $k+1$ can be estimated by aggregating CT information of individual $i$ over the past 8 days, namely,
$\widehat{\Pr}^{(k+1)}\!(X_{i}\!=\!1) = 1 \!-\! \prod_{d=k-7}^k{\prod_{j=1}^n{{ \!\big( 1-\widehat{p}^{(d)}_{i,j} \big) }}}$,
where $\widehat{p}_{i,j}^{(d)}$ is the estimated probability of infection of individual $i$ due to 
contact with individual $j$.
This probability, $\widehat{p}_{i,j}^{(d)}$, can be determined by the CT information ($\tau_{ij}^{(d)}, d_{ij}^{(d)})$, as well as their infection status as follows:
\vspace*{-2mm}
\begin{equation}
\widehat{p}_{i,j}^{(d)}=\exp\left(-\big(\lambda \, \tau_{ij}^{(d)} \,  d_{ij}^{(d)} \, \Psi_{ij}^{(d)}+\epsilon\big)^{-1}\right),
\label{eq:est_pij}
\vspace*{-1mm}
\end{equation}
where $\Psi_{ij}^{(d)} =
1 - \widehat{\Pr}^{(d)}\!(X_i\!=\!0) \, \widehat{\Pr}^{(d)}\!(X_j\!=\!0){\color{orange}}$,
$\lambda$ is a Poisson rate parameter,
and $\epsilon$ is used to avoid division by zero.
Note that \(\widehat{p}_{i,j}^{(t)}\) depends on $\lambda$, which is unknown in practice. We estimate it using a plug-in approach by Ma et al.~\cite{dror_plugin}. More details are given in Sec.~1.2 of the supplemental document.

\noindent{}{\bf Algorithms for multiplicative noise.}
For model \textbf{M2}, recall that $\boldsymbol{x}$ and $\boldsymbol{y}$ represent viral loads of individual samples and pools,
respectively. The core algorithm presented in
\cite{Ghosh2020} uses the well-known \lasso{} estimator, 
$\boldsymbol{\widehat{x}}^{\lasso{}} =\text{arg}\min_{\boldsymbol{x}} \|\boldsymbol{y}-\boldsymbol{Ax}\|^2_2 + \rho \|\boldsymbol{x}\|_1$ \cite{THW2015}, where $\rho$ is a smoothness parameter. 
\lasso{} exploits the sparsity of $\boldsymbol{x}$ but uses no SI. Despite the multiplicative nature of the noise, 
\lasso{} yields good estimation performance \cite{Ghosh2020} in terms of three commonly used measures: 
({\em i}) {\em relative root mean squared error} (RRMSE) $= \|\boldsymbol{x}-\boldsymbol{\widehat{x}}\|_2/ \|\boldsymbol{x}\|_2$;
({\em ii}) {\em false negative rate} (FNR) = $\#$incorrectly detected negatives$\big/ \#$true positives; and
({\em iii}) {\em false positive rate} (FPR) = $\#$incorrectly detected positives$\big/ \#$true negatives.
Note that FNR $=1-$ sensitivity and FPR $=1-$ specificity.

In some cases, the $n$ individuals in $\boldsymbol{x}$ can be partitioned into $n_1 \ll n$ disjoint 
groups of people, for example family members, 
who interact closely with each other and are thus likely to pass the virus between group members. 
This family-style structure leads to a situation where either all members of the group are uninfected, 
or a majority of members are infected. Note that the family-style structure also includes groups of coworkers, 
students taking a course together, and people sharing common accommodation. 
If reliable SI about how the $n$ individuals are partitioned into families is available, and only 
a small portion of families, $n_2 \ll n_1$, are infected, then \lasso{} can be replaced by 
{\em group square-root \lasso{}} (\sqrtglasso{}) \cite{Yuan2006}.\footnote{
We observed that \sqrtglasso{}, which has an $\ell_2$ data fidelity term instead of a squared $\ell_2$ one \cite{Belloni2011},
outperformed \textsc{Glasso}. In contrast, conventional \lasso{} outperformed \textsc{Sqrt-Lasso}.} 
The latter is defined as 
\vspace{-3mm}
\begin{equation}
\boldsymbol{\widehat{x}}^\sqrtglasso{} = \operatorname{arg}\min_{\boldsymbol{x}} \|\boldsymbol{y}-\boldsymbol{Ax}\|_2 + \rho \sum_{g=1}^{n_1}\|\boldsymbol{x}_g\|_2,
\vspace{-2mm}
\end{equation}
where $\boldsymbol{x}_g$
consists of viral loads of people from the $g$th family.

\begin{figure*}[!t]
    \vspace{-1mm}
	\begin{tabular}{@{}cc@{}c@{}c@{}c@{}}
	\multicolumn{2}{l}{\underline{Sparsity:} \hspace{8mm} \underline{$2.12\%$}} & \underline{$3.98\%$} & \underline{$6.01\%$} & \underline{$8.86\%$}  \vspace{0mm} \\
  \rotatebox[origin=l]{90}{\hspace{14mm}\underline{\textbf{M1}}} & 
		\includegraphics[width=0.24\linewidth]{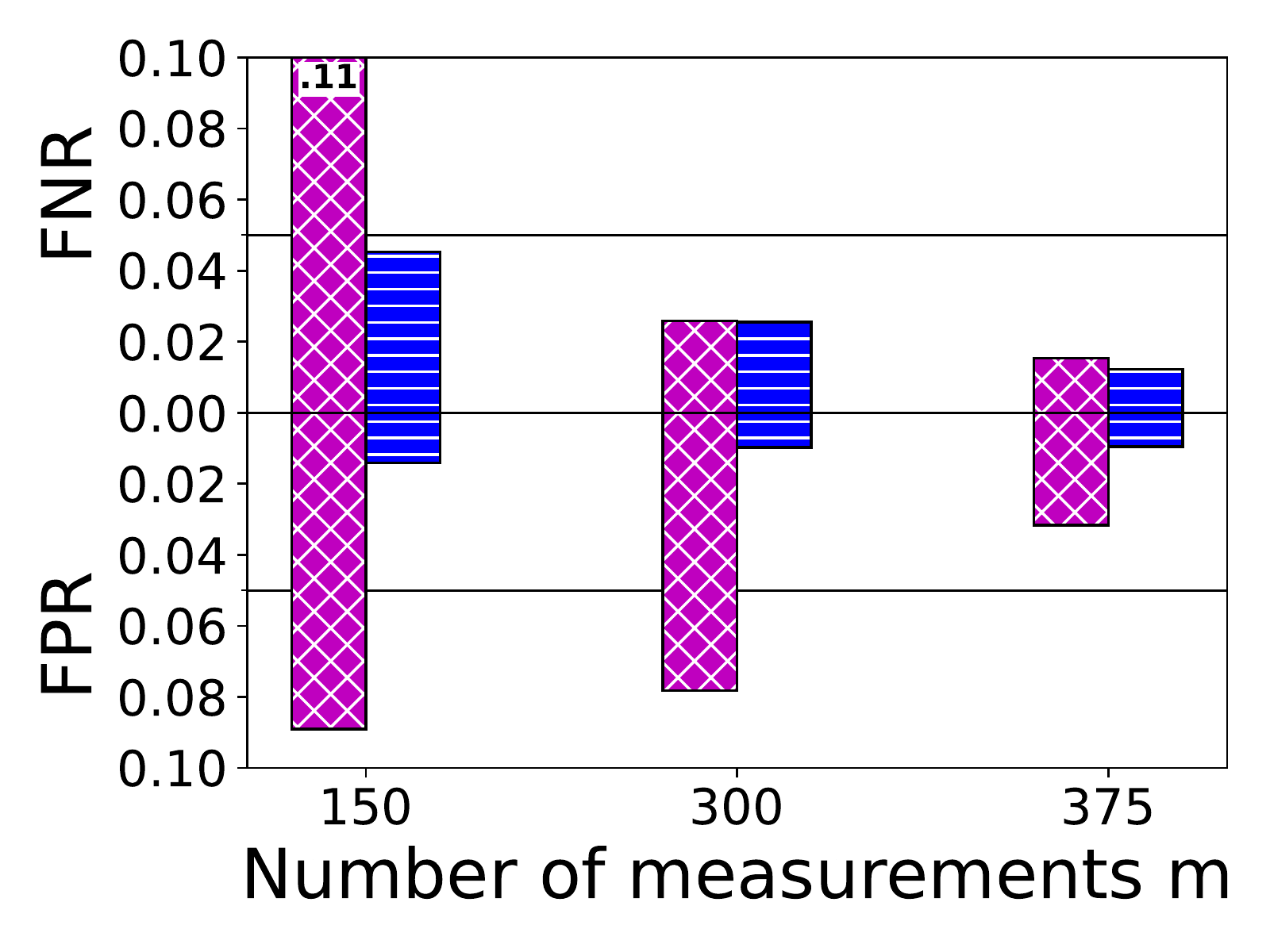} &
		\includegraphics[width=0.24\linewidth]{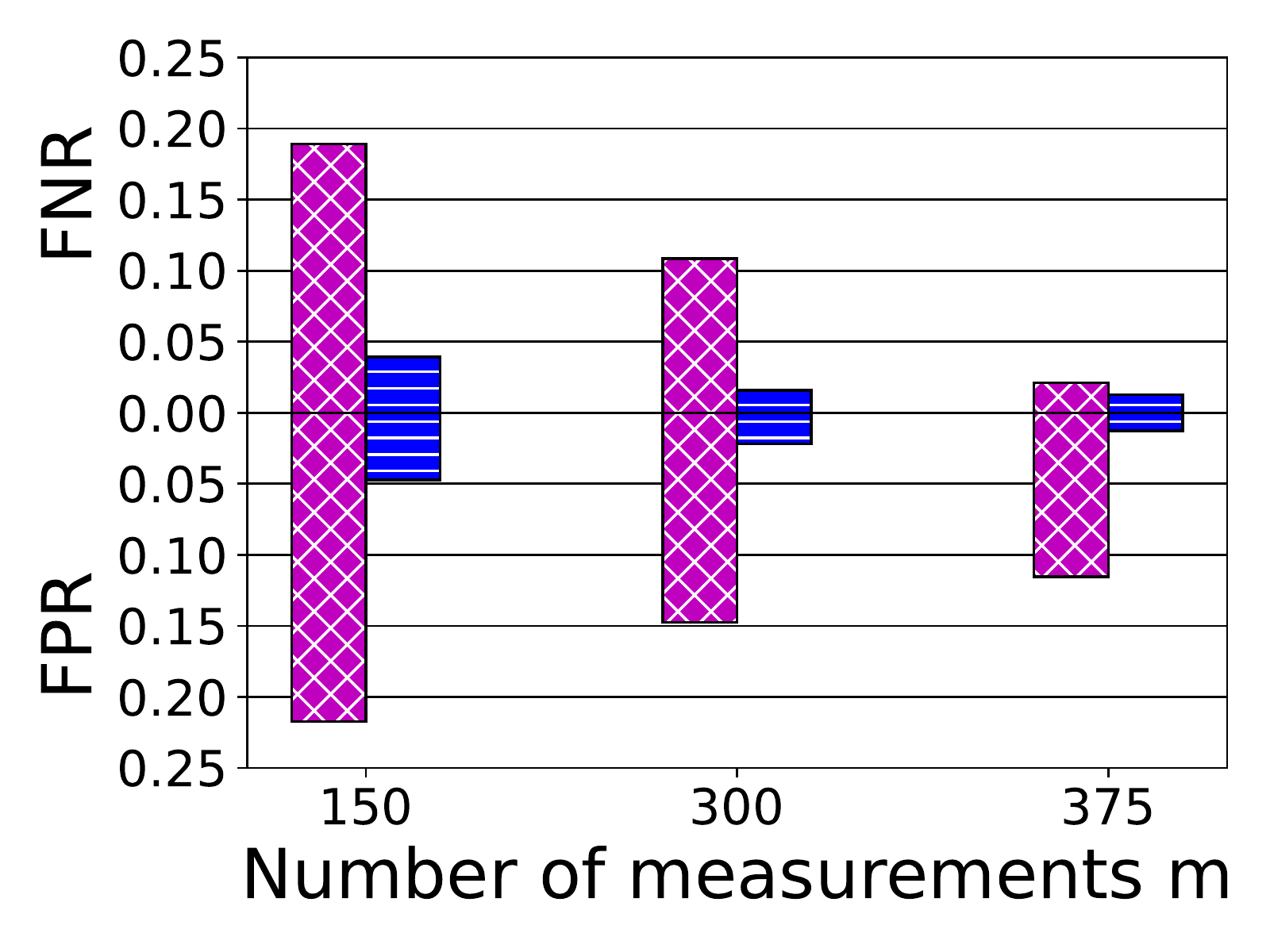} &
		\includegraphics[width=0.24\linewidth]{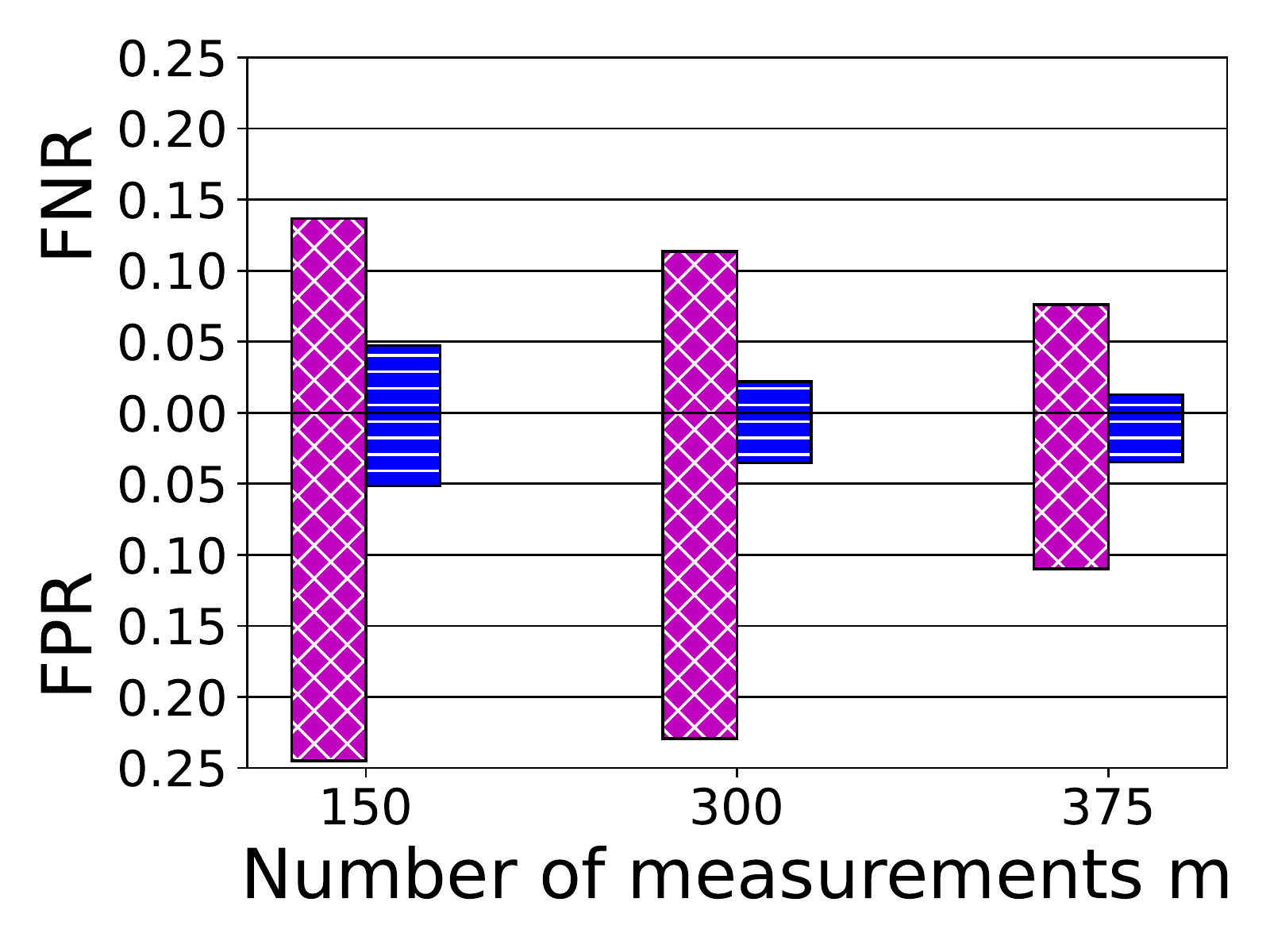} &
		\includegraphics[width=0.24\linewidth]{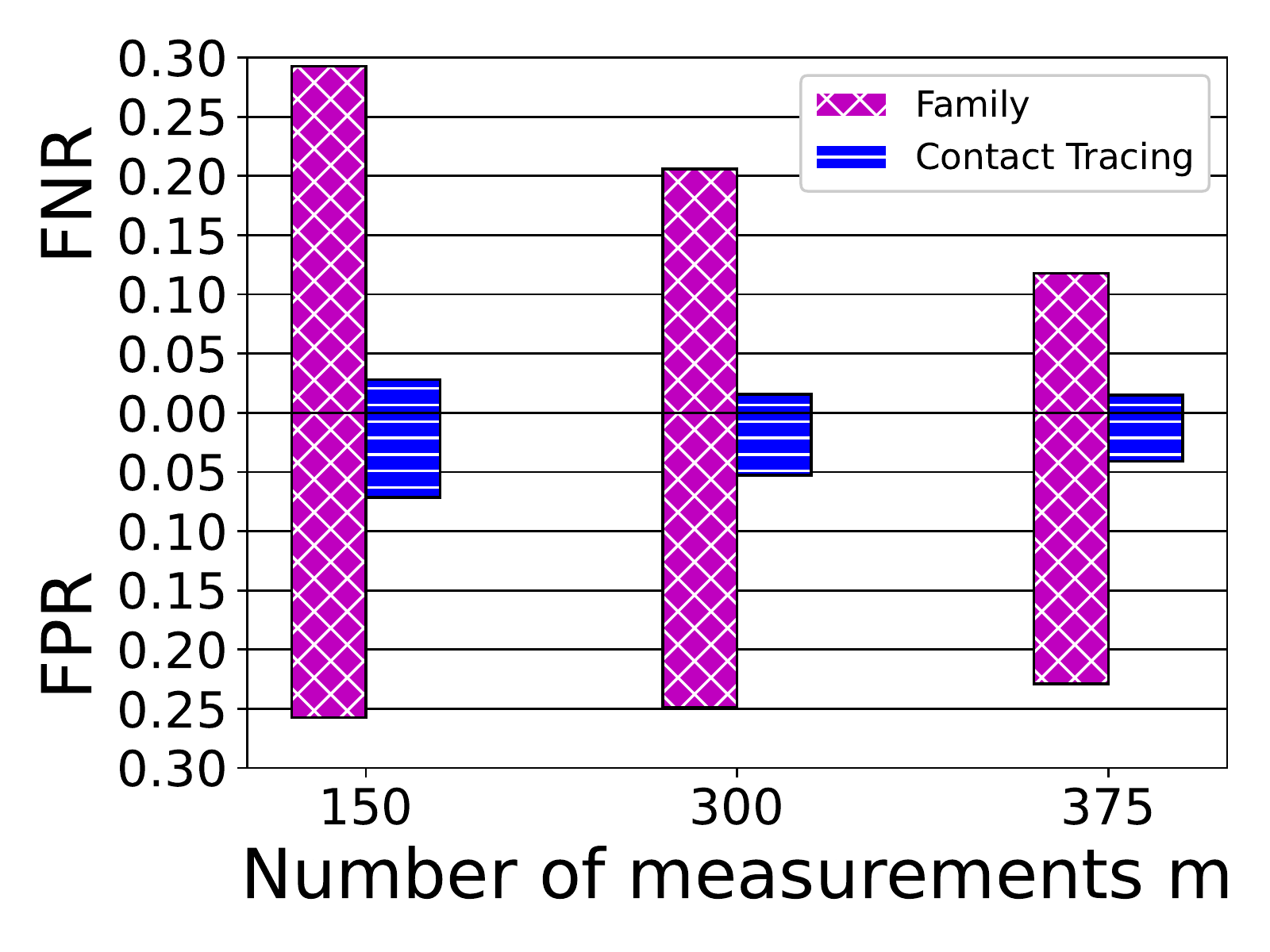} \vspace{-2mm} \\ 
		\rotatebox[origin=l]{90}{\hspace{14mm}\underline{\textbf{M2}}} & 
		\includegraphics[width=0.24\linewidth]{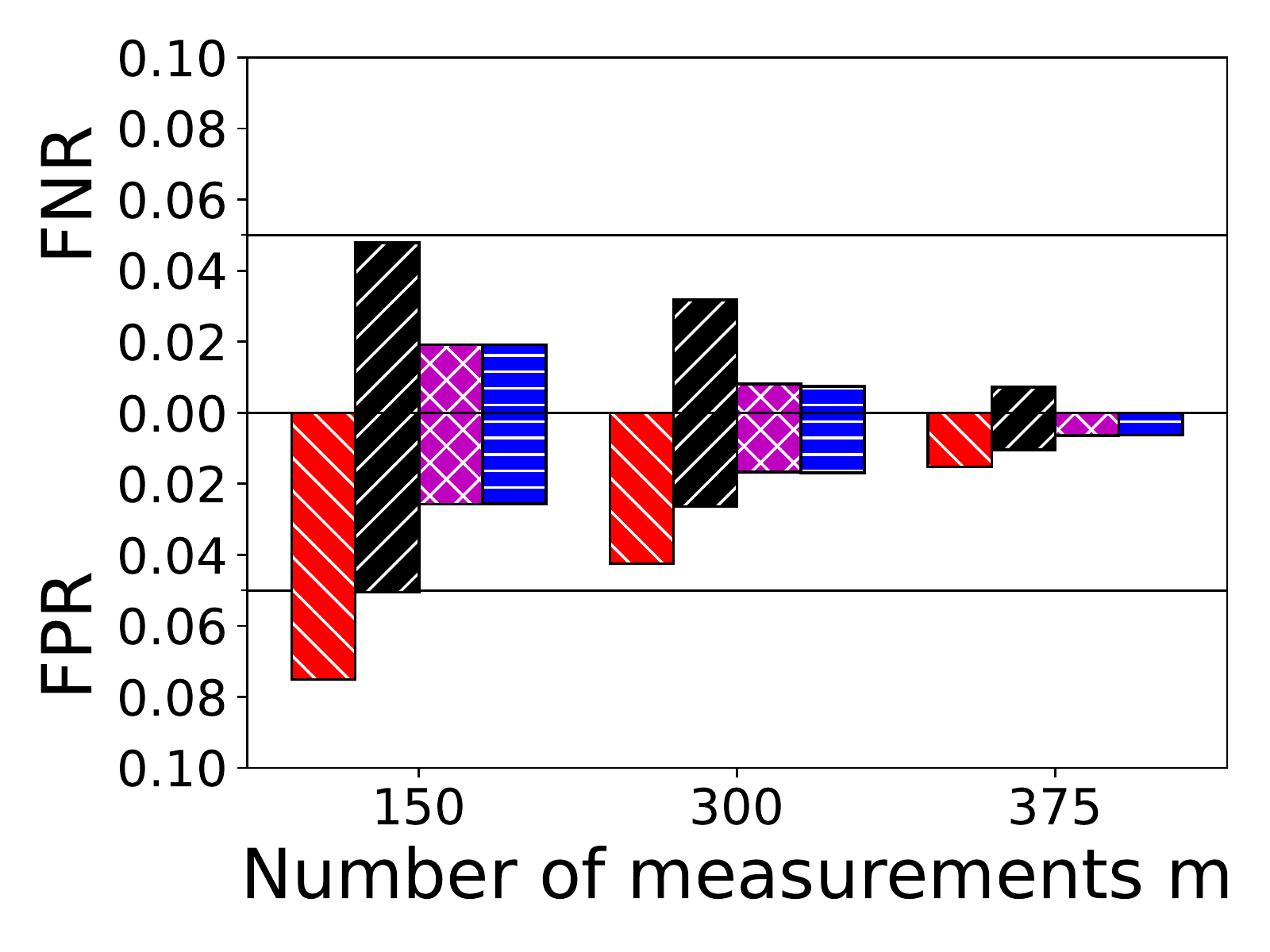} &
		\includegraphics[width=0.24\linewidth]{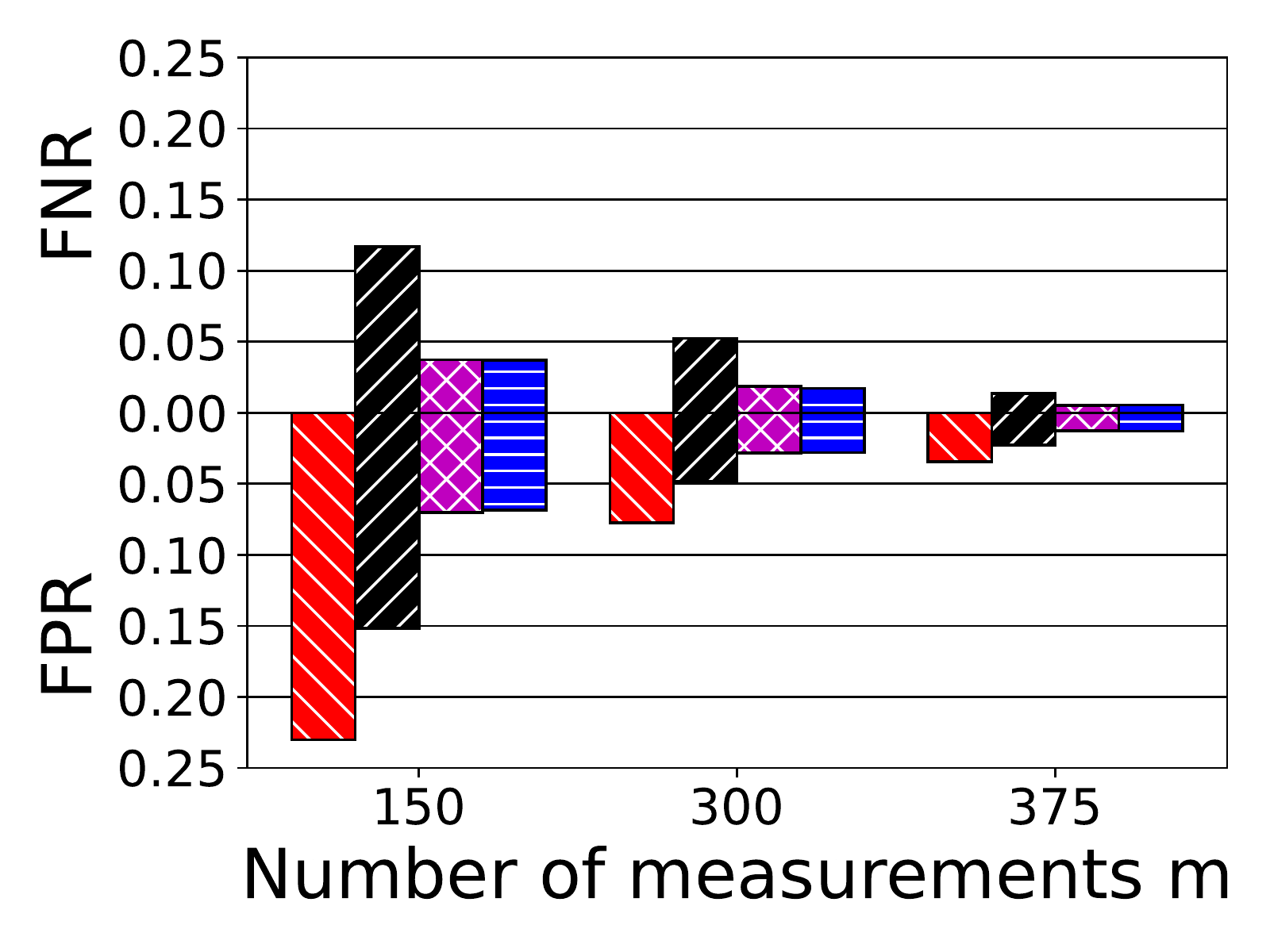} &
		\includegraphics[width=0.24\linewidth]{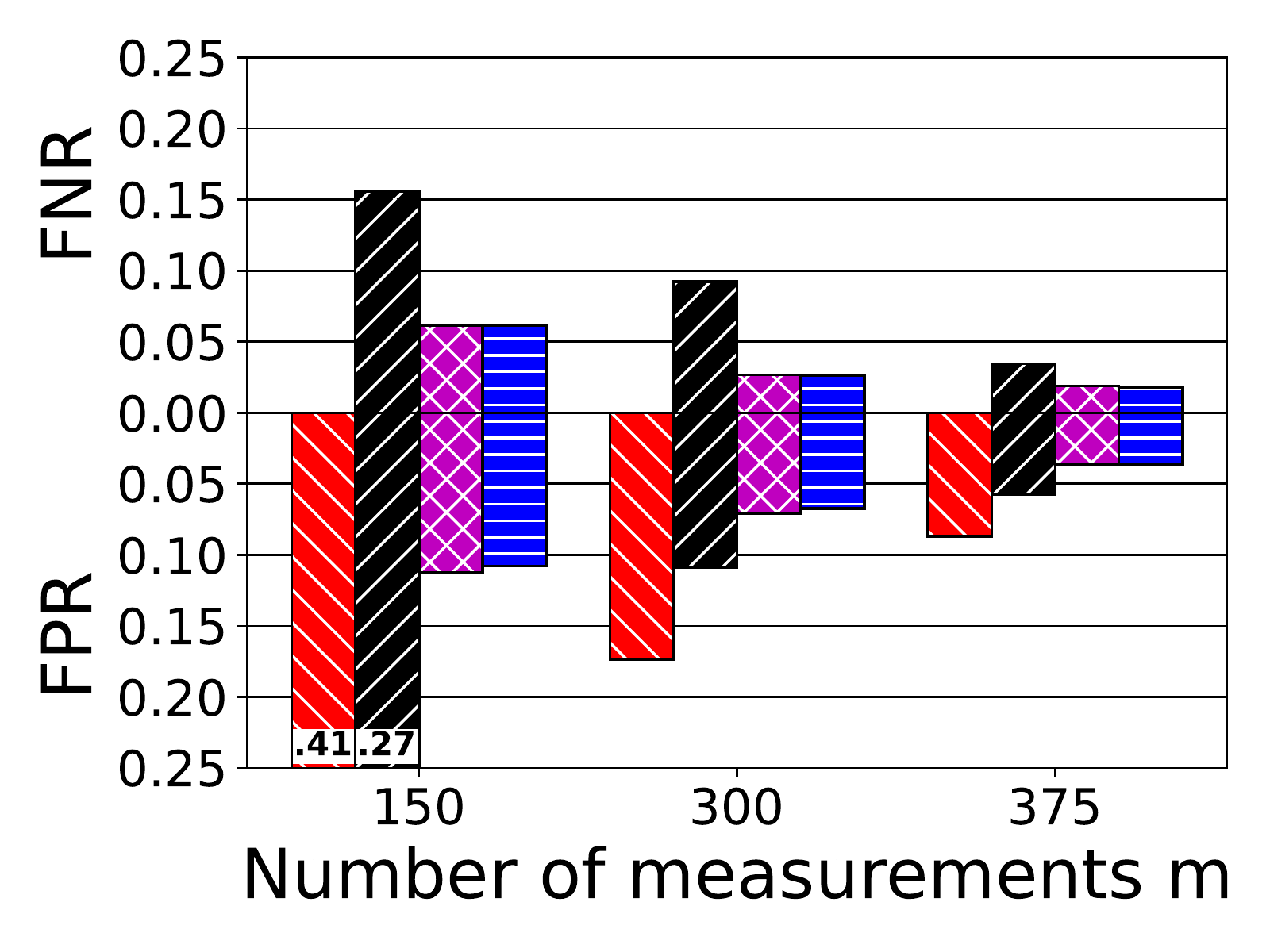} &
		\includegraphics[width=0.24\linewidth]{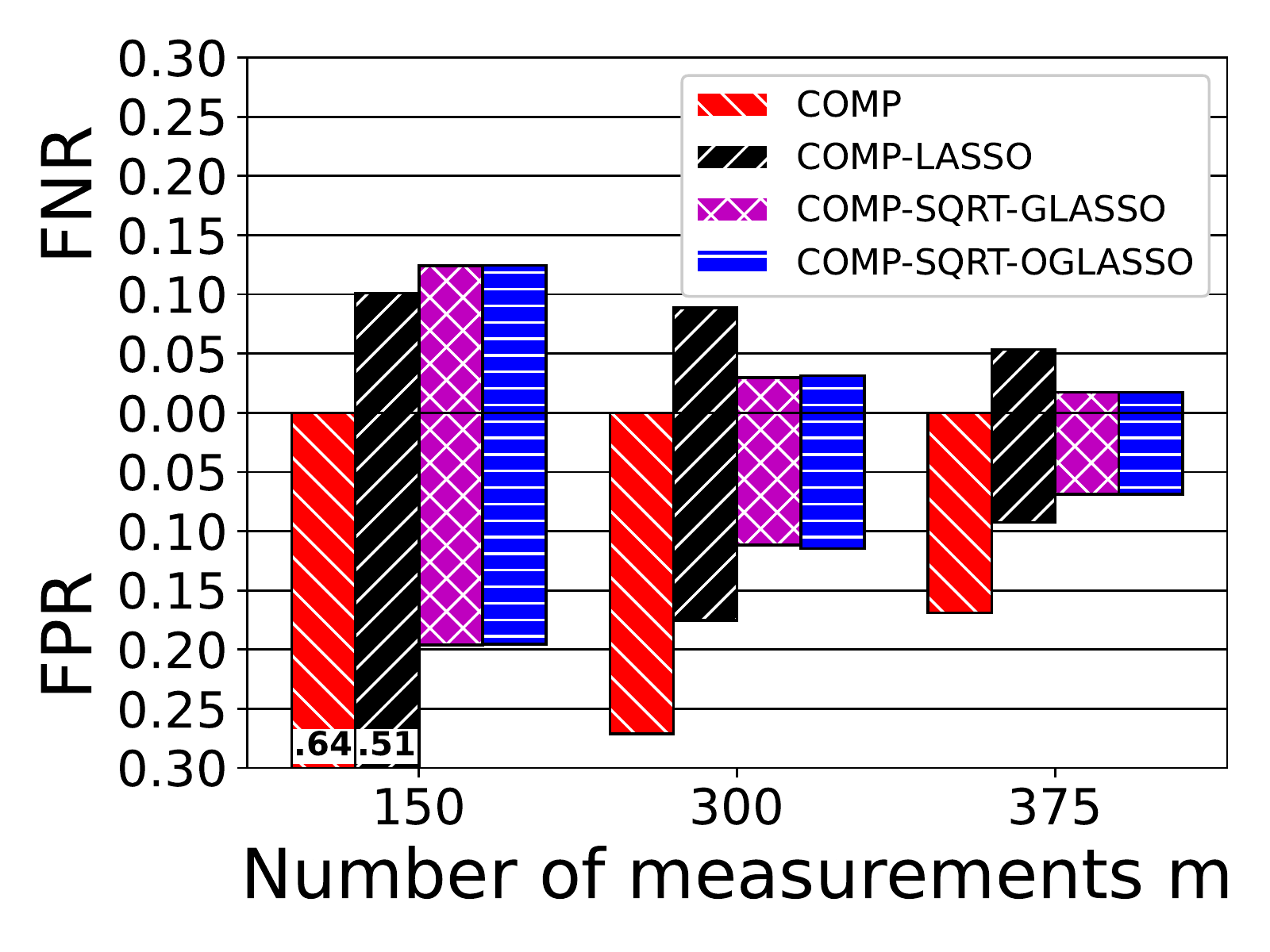}
	\end{tabular}
  \vspace{-6mm}
  \caption{Performance of the proposed group testing methods \textbf{M1} (top row) with binary noise and \textbf{M2} (bottom row) with multiplicative noise at four averaged sparsity levels and three measurement levels for a population of $n = 1000$ individuals.}
  \label{fig:res-m2}
  \vspace{-4mm}
\end{figure*}

In other cases, accurate family SI may be unavailable or unreliable. 
Moreover, family SI does not account for contact between members of different families. 
In such cases, we use CT SI commonly available via Bluetooth~\cite{Hekmati2020}
to directly \emph{infer} family-style structure using clique detection algorithms;
contacts between members of different families can also be considered to be small cliques. 
In particular, we use the Bron--Kerbosch algorithm \cite{Bron1973} to find maximal cliques in the CT graph, and label each clique as a family. Note that one could generate these groups differently~\cite[Sec.~7]{Jacob2009}, 
for example, decomposition into $k$-clique communities \cite{Palla2005}. 
However, such a decomposition may partition the $n$ individuals into $n_3 \ll n$ family structures that \emph{overlap} with each other, unlike 
the earlier case of \emph{disjoint} families.
In a scenario with overlapping families, we use the {\em overlapping group square-root \lasso{}} (\compsqrtoglasso{})
estimator \cite{Jacob2009},
\vspace{-2mm}
\begin{equation}
\boldsymbol{\widehat{x}}^\sqrtoglasso{} = \operatorname{arg}\min_{\boldsymbol{x}} \|\boldsymbol{y}-\boldsymbol{Ax}\|_2 + \rho\, \Omega_{\text{overlap}}(\boldsymbol{x}),
\vspace{-2mm}
\end{equation}
where $\Omega_{\text{overlap}}(\boldsymbol{x}) = \inf_{\boldsymbol{v} \in \mathcal{V}_G, \sum_{g \in G} \boldsymbol{v}_g = \boldsymbol{x}}  \sum_{g \in G} \|\boldsymbol{v_g}\|_2$,
$G$ denotes a set of possibly overlapping groups each containing a subset of the $n$ individuals in $\boldsymbol{x}$, 
$\mathcal{V}_G$ is a set of $|G|$-tuples of vectors $\boldsymbol{v} = (\boldsymbol{v}_g)_{g \in G}$, 
$g$ is an index for the groups in $G$, 
and $\boldsymbol{v}_g \in \mathbb{R}^n$ is a vector whose support is a subset of $g$. Advantages of \textsc{Oglasso} over \textsc{Glasso} for overlapping groups are summarized in Fig.~1 and Sec.~3 of \cite{Jacob2009}.

In all three algorithms, we imposed a non-negativity constraint on $\boldsymbol{x}$. 
Moreover, all algorithms were preceded by a step that executed {\em combinatorial orthogonal matching pursuit} (\comp{}). 
\comp{} declares all samples that contributed to a pool with zero viral load to be negative.
This \comp{} preprocessing step reduces the problem size and improves all three algorithms' performance, as well as speed. 
We henceforth refer to our algorithms as \complasso{}, \compsqrtglasso{} and \compsqrtoglasso{}.

%=====

\vspace{-2mm}
\section{Numerical Results}
\label{sec:results}
\vspace{-1mm}

We now present numerical results obtained for the two models, \textbf{M1} and \textbf{M2}.
The data was generated based on Sec.~\ref{sec:data_gen},
and group testing inference was performed using the algorithms proposed in Sec.~\ref{sec:algos}.
We generated datasets using four levels of cross-clique contacts, leading to four averaged sparsity levels, $2.12\%$, $3.98\%$, $6.01\%$, and $8.86\%$, for $\boldsymbol{x}$.
At each sparsity level, we perform pooling experiments using Kirkman triple matrices 
as proposed in \cite{Ghosh2020} for three measurement setups, $m \in \{150, 300, 375\}$.
Measurement vectors $\y$ for M1 were generated using probabilities for erroneous binary tests, $\Pr(y_i=1|w_i=0)=0.001$ and $\Pr(y_i=0|w_i>0)=0.02$, per Hanel and Thurner~\cite{hanel2020boosting}.
Vectors $\y$ for M2 were generated by setting the parameter reflecting the strength of noise in RT-PCR to $\sigma^2 = 0.01$.
Fig.~\ref{fig:res-m2} shows the performance of the proposed algorithms in terms of FNR and FPR%
\footnote{For M1, we chose to report the FPR and FNR pair such that the sum of the two error rates is minimized. The complete ROC curves are shown in Sec.~2.1 of the supplemental document.
For M2, we reported the error rates by thresholding the estimated viral load using $\tau = 0.2$. We noticed that the error rates do not change much when $\tau$ 
varies between $0$ and $1$.
}
averaged across the inference results obtained for the time window of $50$ days described in Sec.~\ref{sec:data_gen}.

For model \textbf{M1}, we tested the family denoiser~\eqref{eq:denoiser_f} and the CT denoiser~\eqref{eq:denoiser_ct}.
Fig.~\ref{fig:res-m2} shows that the CT denoiser outperforms the family denoiser in all settings.
Both algorithms yield lower (better) FNR and FPR as the number of measurements, $m$, increases.
Moreover, the CT denoiser's error rates are below
$0.05$, except for the challenging cases where the sparsity level is 
$8.86\%$ and $m\in\{150,300\}$.

For model \textbf{M2}, we tested four algorithms: \comp{}, \complasso{}, \compsqrtglasso{}, and \compsqrtoglasso{}. 
The results show that both \compsqrtglasso{} and \compsqrtoglasso{} outperform \complasso{} in terms of FNR and FPR, 
which shows the benefit of using CT SI. 
Note that \compsqrtoglasso{} performs on par with \compsqrtglasso{}, 
even though the former infers everything on the fly from CT SI without explicit access to family SI.
The \comp{} algorithm by itself produces no false negatives (corresponding to $\text{FNR} = 0$), but many false positives. 
Further, all four algorithms yield lower (better) FNR and FPR as $m$ increases or the averaged sparsity level decreases. Finally, we remark that \compsqrtoglasso{} outperforms \compsqrtglasso{} for more general CT graphs consisting of slightly 
incomplete cliques. We refer the readers to Sec.~3 of the supplemental document for details.

Our algorithms, when presented with SI, 
reduce the FNR and FPR, 
despite not knowing which individuals are infected within each infected family (around $70\%$ of the individuals are infected within an infected family on average).
Note that none of the algorithms for model \textbf{M2} make use of previous inference results, 
whereas the CT denoiser for model \textbf{M1} uses such information.
This distinction makes the two approaches applicable in different scenarios, namely, the CT denoiser can be used for 
a CT and testing program where the same population is tested at regular intervals, e.g., warehouse employees,
whereas \compsqrtoglasso{} is useful when a population has to be tested only once. 
Furthermore, while model \textbf{M1} performs well in the presence of erroneous binary tests, it does not yield viral load estimates as \compsqrtoglasso{} does.
Viral load estimates could prove to be useful, since there is a positive correlation between mortality 
and viral loads~\cite{Westblade2020,Pujadas2020}.

%=====

\vspace*{2mm}
\noindent{}{\bf Acknowledgment:}\ \ The authors thank Junan Zhu for allowing them to use his implementation of 
GAMP with SI in their implementation of the family and CT denoisers.

\small
\bibliographystyle{IEEEbib}
\bibliography{refs}
\normalsize

\setcounter{section}{0}
\newpage

\placetextbox{0.5}{0.97}{\fbox{\parbox{7in}{\textbf{Note: This is a supplemental document for ``Contact tracing enhances the efficiency of COVID-19 group testing,'' 
submitted to \emph{2021 IEEE International Conference on Acoustics, Speech and Signal Processing} by Ritesh Goenka,$^{\star}$ Shu-Jie Cao,$^{\star}$ Chau-Wai Wong, Ajit Rajwade, and Dror Baron. RG and SJC have made equal contributions to the paper.
}}}}

\section{Algorithmic Details of M1}
\subsection{Family Denoiser}
We now formalize a family-based infection mechanism that can be used in
designing group testing algorithms for improving the detection accuracy.
We define $\M_{\F}$ as the set of indices of all members of family $\F$.
We say that $\F$ is {\em viral} when there exists viral material in the family.
Next, define the infection probability of individual $i$ within viral family 
$\F$ as $\pip = \Pr(X_i = 1 \mid \Fviral)$, for all $i \in \M_{\F}$, and $\pif = \Pr(\Fviral)$.
Note that the infection 
status of individuals in a viral family are conditionally 
independent and identically distributed (i.i.d.).

Under our definition, family $\F$ being viral need not be attributed to any individual $i \in \M_{\F}$.
After all, viral material can be on an infected pet or contaminated surface. 
For this model, once the family is viral, the virus
spreads independently with a fixed probability $\pip$.
Of course, our simplified model may not accurately reflect reality. 
That said, without a consensus in the literature on how coronavirus spreads, 
it is unrealistic to create a more accurate model.
On the other hand, our model is plausible, and we will see that it is mathematically tractable.

We further assume that individuals cannot be infected unless the family is viral,
i.e., $\Pr(X_i = 1 \mid \F \text{ not viral}) = 0$.
The family structure serves as SI and allows the group testing algorithm to impose the constraint that people living together have strongly correlated health status.

Next, we derive the exact form of the denoiser \eqref{input_d} by incorporating the family-based infection mechanism.
Denote the pseudodata of the members of family $\F$ as $\v_{\F} = (v_i)_{i \in \M_{\F}}$, 
the family-based denoiser for $i$th individual can be decomposed as follows:
\begin{subequations}
\begin{align}
&g_{\text{in}}^\text{family}(\v_\F) \notag \\
=& \E \left[ X_{i} \mid \v_{\F} \right] \\
=& \Pr(X_i = 1 \mid \v_{\F}) \\ 
=& \Pr(X_i = 1, \Fviral \mid \v_{\F})  \\
=& \Pr(\Fviral \mid \v_{\F}) \Pr(X_i = 1 \mid \v_{\F},\Fviral),
\label{f_den}
\end{align}
\end{subequations}
where the first term of \eqref{f_den} is
\begin{align}
&\Pr(\Fviral \mid \v_{\F}) \nonumber \\
=& \, \frac{f(\v_{\F},\Fviral)}{f(\v_{\F},\Fviral) + f(\v_{\F},\F\text{ not viral}) }.
\label{p_infect}
\end{align}
The two quantities in \eqref{p_infect} can be further expanded as
\begin{subequations}
\begin{align}
&f(\v_{\F},\F\text{ not viral}) \\
= &(1-\pih) \, f(\v_{\F} \mid \F\text{ not viral}) \\
= &(1-\pih) \prod_{i\in \M_{\F}}
\normaldensity{v_i}{0}{\Delta},
\end{align}
\end{subequations}
and
\begin{subequations}
\begin{align}
&f(\v_{\F},\ \Fviral) = \pih \, f(\v_{\F} \mid \Fviral)\\
\begin{split}
  =\, &\pih\sum_{\x_k \in \Omega_{\F}} \prod_{i \in \M_{\F}} \\ 
  & \Big[ 
  f(v_{i}|X_{i}=x_{k,i})
  \Pr( X_i = x_{k,i} | \Fviral ) 
  \Big],
\end{split}
\label{sum over prod}
\normalsize
\end{align}
\end{subequations}
where $\normaldensity{x}{\mu}{\sigma^2} := \frac{1}{\sqrt{2 \pi \sigma^2}} \exp \left(  \frac{(x-\mu)^2}{2 \sigma^2} \right)$, 
and $\Omega_{\F} = \{0...00,\  0...10,\ \dots,\ 1...11\}$ is a power set 
comprised of $2^{|\M_{\F}|}$ distinct infection patterns for family $\F$.
The second term of \eqref{f_den} can be simplified as follows:
\begin{subequations}
\begin{align}
\notag
&\Pr(X_i = 1 \mid \v_{\F},\Fviral) \\
= & \Pr(X_i = 1 \mid v_i,\Fviral)  \\
=&\Pr(X_i = 1, v_i \mid \Fviral) \, / \, \Pr(v_i \mid \Fviral) \\
=&\dfrac{\pip \, \normaldensity{v_i}{1}{\Delta}}
{\pip \, \normaldensity{v_i}{1}{\Delta} +  \left(1-\pip\right) \, \normaldensity{v_i}{0}{\Delta}} \\
=&  \left( 1 + \frac{1-\pip}{\pip} \cdot \frac{\normaldensity{v_i}{0}{\Delta}}{\normaldensity{v_i}{1}{\Delta}} \right)^{-1} \\
=& \left( 1 + \left( \pip^{-1} - 1 \right)
  \exp \Big[ \big(v_i-\tfrac{1}{2}\big) \big/ \Delta \Big] \right)^{-1}.
\end{align}
\end{subequations}

\subsection{Contact Tracing Denoiser}
While family structure SI characterizes part of the spread of the disease,
individual members of a family will presumably all come in close contact with each other,
hence CT SI will include cliques for these individuals. Additionally, CT SI
describes inter-family contacts. Therefore,
CT SI can characterize the spread of the disease more comprehensively than family SI.
To exploit the CT SI, we encode it for each individual $i$ into the prior probability of infection, $\Pr(X_i=1)$,
and use the following scalar denoiser: 
\begin{subequations}
\begin{align}
&g_{\text{in}}^\text{CT}(v_i) \nonumber \\
=& \E \left[ X_i|v_i \right] = \Pr(X_i = 1|v_i) \\
=& f(v_i|X_i=1)\Pr(X_i=1)/f(v_i) \\
=& \left\{ 1 \!+\! \big[\Pr(X_i\!=\!1)^{-1} \! - \! 1\big]
  \exp \Big[ \big(v_i-\tfrac{1}{2}\big) \big/ \Delta \Big] \!\right\}^{-1}.
\end{align}
\end{subequations}
Here, $\Pr(X_i\!=\!1)$ for day $k+1$ can be estimated by aggregating CT information of individual $i$ over a so-called \emph{SI period} from day $k-7$ to day $k$ as follows
\begin{equation}
\widehat{\Pr}^{(k+1)}(X_{i}=1) = 1 - \prod_{d=k-7}^k{\prod_{j=1}^n{{ \!\big( 1-\widehat{p}^{(d)}_{i,j} \big) }}},
\label{eq:SI_period_aggre}
\end{equation}
where $\widehat{p}_{i,j}^{(d)}$ is the estimated probability of infection of individual $i$ due to 
contact with individual $j$.
This probability, $\widehat{p}_{i,j}^{(d)}$, can be determined by the CT information ($\tau_{ij}^{(d)}, d_{ij}^{(d)})$, as well as their infection status as follows:
\begin{equation}
\widehat{p}_{i,j}^{(d)}=\exp\left(-\big(\lambda \, \tau_{ij}^{(d)} \,  d_{ij}^{(d)} \, \Psi_{ij}^{(d)}+\epsilon\big)^{-1}\right),
\label{eq:est_pij_2nd}
\end{equation}
where $\Psi_{ij}^{(d)} =
1 - \widehat{\Pr}^{(d)}\!(X_i\!=\!0) \, \widehat{\Pr}^{(d)}\!(X_j\!=\!0){\color{orange}}$,
$\lambda$ is an unknown Poisson rate parameter,
and $\epsilon$ is used to avoid division by zero. 
We estimate $\lambda$ with maximum likelihood (ML) using the pseudodata of all individuals, i.e.,
\begin{equation}
\widehat{\lambda}^{\text{ML}} = \arg\max_{\lambda}\ \prod_{i=1}^n f(v_i|\lambda)
\label{eq:mle},
\end{equation}
where \(f(v_i|\lambda) = f(v_i|X_{i}=1) \,  \Pr(X_{i}=1|\lambda)+f(v_i|X_{i}=0) \, \Pr(X_{i}=0|\lambda)\).
Once $\widehat{\lambda}^{\text{ML}}$ is obtained, it is plugged into \eqref{eq:est_pij_2nd} for calculating the prior probability\cite{dror_plugin}.
Note that this plug-in strategy is also used for two other denoisers, namely, $\lambda=\rho$ for $g_{\text{in}}^\text{Bernoulli}(v_i)$ and $\lambda = (\pih, \pip)$ for $g_{\text{in}}^\text{family}(\v)$.

\section{Additional Results for M1}

\begin{figure*}[!t]
\centering
\includegraphics[scale=0.62]{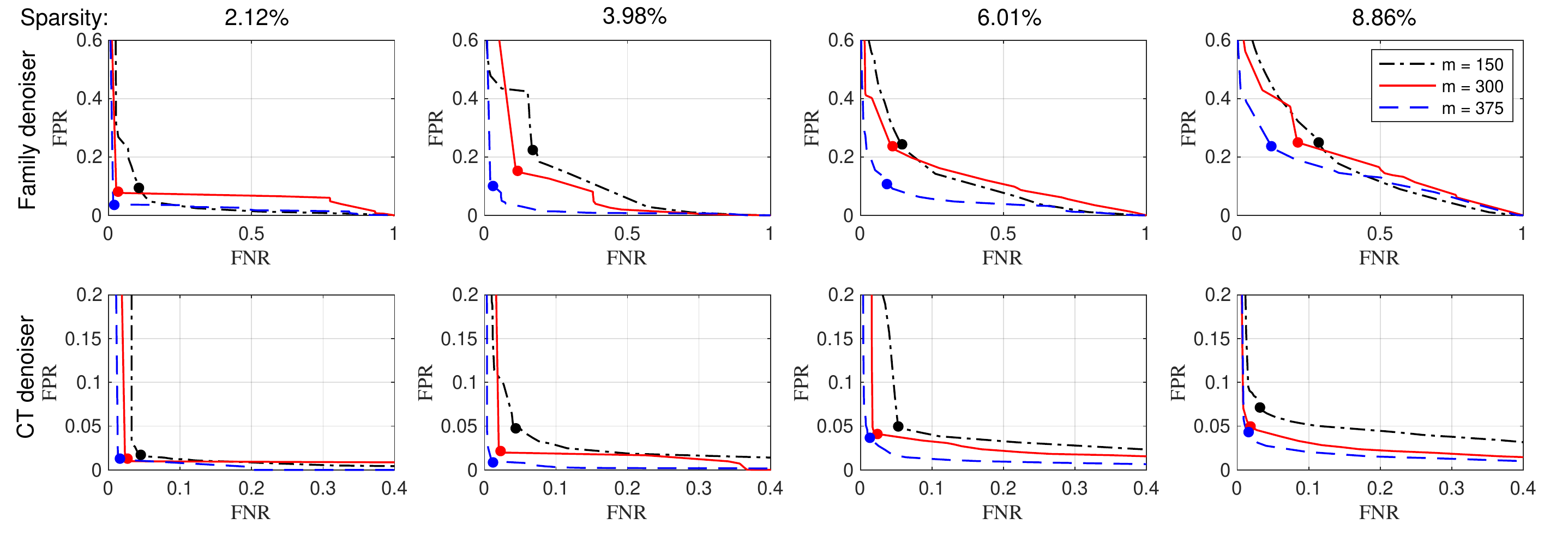}
\caption{Performance of \textbf{M1} in terms of ROC when family denoiser (top row) and CT denoiser (bottom row) are used.
Columns correspond to averaged sparsity levels ranging from $2.12\%$ to $8.86\%$.
Within each plot, the performance under three measurement levels for a population of $n = 1000$ individuals is compared. 
The dot on each curve corresponds to an operating point that minimizes the sum of FPR and FNR.
The CT denoiser significantly outperforms the family denoiser with error rates mostly below $0.05$.
The estimation problem is more challenging when fewer measurements are used at a higher sparsity level.
}
\label{fig:roc_M1}
\end{figure*}

In Sec.~4 of the main paper, we reported the performance of M1 in a compact way, due to space limitations, 
by choosing a representative operating point on an ROC curve instead of using the whole curve.
In this section, we provide complete ROC curves that correspond to the top row of Fig.~3 of the main paper.
Fig.~\ref{fig:roc_M1} illustrates M1's performance for family and CT denoisers at different measurement and sparsity levels.
The dot on each curve corresponds to the operating point that minimizes the total error rate, i.e., the sum of FPR and FNR, as reported in Sec.~4 of the main paper.
The closer a dot is to the origin of the FPR--FNR plane, the better the performance it reflects.
Comparing the ROC curves in the top row to those in the bottom row, we note that the CT denoiser significantly outperforms the family denoiser at all sparsity levels.
The CT denoiser, with most of its FNR and FPR $< 5\%$, can achieve as low as $15\%$ of the total error rate of the family denoiser.
Across different sparsity levels, the algorithm performs less accurately as 
the sparsity level increases. 
In each plot, lower measurement rates make it
more challenging for the group testing algorithm.

We also examine the stability of the thresholds corresponding to the operating points we selected 
to report results in Fig.~3 of the main paper. 
Our empirical results reveal that at a particular sparsity level, the variation of the threshold due to different design matrices or denoisers is less than $0.003$.
As the sparsity level increases from $2.12\%$ to $8.86\%$, the threshold only drops
from $0.160$ to $0.137$. Hence, the threshold for minimizing the total error rate is insensitive to the testing conditions.

\section{Additional Experiments for M1}
\subsection{Using Prior Knowledge of the Infection Status}
In this subsection, we examine the advantage that prior knowledge of the population's infection status in the startup phase
provides our proposed algorithm for the M1 binary model.
As stated in Sec.~3 of the main paper and in \eqref{eq:SI_period_aggre}, 
our algorithm iteratively uses the updated probability of infection, $\widehat{\Pr}(X_i=1)$, 
estimated from an SI period of 8 immediately preceding days. 
Note that for days $k < 8$, 
we had to use the ground-truth infection status of each individual in the startup phase to generate the results reported in Sec.~4 of the main paper.
However, ground-truth infection data from 
the startup phase may provide 
our approach an unfair advantage over the 
algorithms proposed for M2. Below, we investigate whether this advantage is significant.

\begin{figure}[!t]
  \vspace{-4mm}
  \includegraphics[scale=0.33]{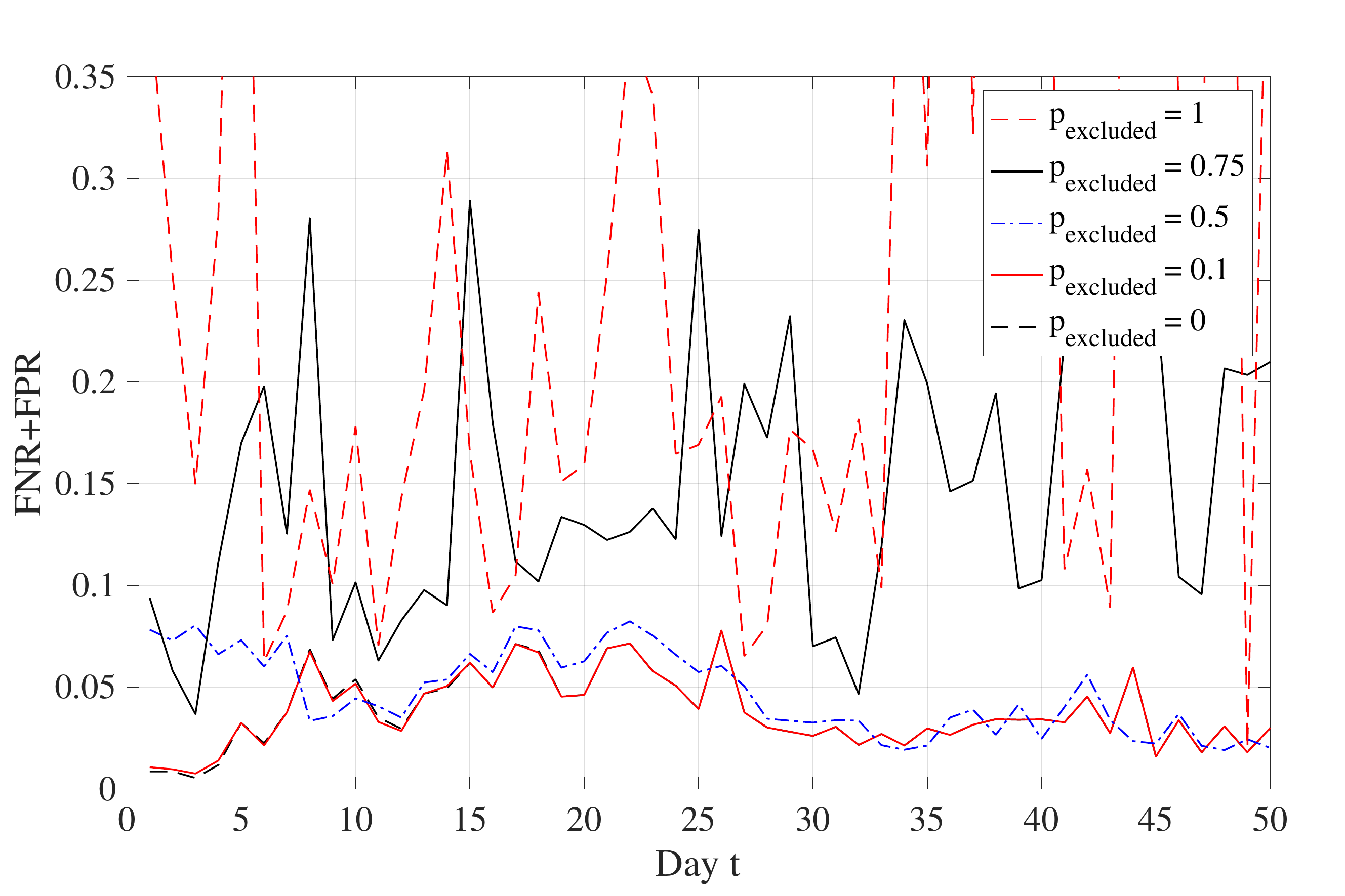}
  \vspace{-8mm}
  \caption{Performance of M1 when a proportion, $p_{\text{excluded}}$, of the population's health states in the startup phase is unknown.
  The curves reveal that in the absence of up to $50\%$ prior knowledge of the infection status of the population, 
  the accuracy of M1 is close to that when complete startup information is available.
  }
  \label{fig:edge_due_to_prior}
\end{figure}

We examine how varying the amount of startup
information impacts our algorithm's quality.
Specifically, we randomly replace a portion, $p_{\text{excluded}} \in \{0, 0.1, 0.5, 0.75, 1\}$, 
of the population's infection status by an estimated probability of infection, e.g., $5\%$, for a setup that has a true averaged sparsity level of $7.2\%$.
Using a probability instead of a binary value, $0$ or $1$, gives the algorithm soft probabilistic information instead of hard ground-truth style information.
Fig.~\ref{fig:edge_due_to_prior} shows that even with 50\% prior knowledge of the infection status of individuals, 
our detection accuracy for M1 is close to that when using complete prior information after ramping up for 8 days.
The averages of the total error rates across time for increasing $p_{\text{excluded}}$ are $0.038$, $0.039$, $0.046$, $0.148$, and $0.407$, respectively. 
We also tried to replace the startup infection status by an estimated probability of infection of $10\%$, but only observed negligible performance differences.
The results show that
the CT algorithm is robust to the absence of up to $50\%$ of startup infection information.

\subsection{Duration of Startup SI Period}
We investigated the impact of the duration of the startup SI on estimation performance.
In principle, the longer the SI duration, the more accurate we expect the results to be.
There is a trade-off between the accuracy of our algorithm and the startup SI infection status information that needs to be pre-collected before the initialization of the testing algorithm.
In our experiment, we tested three startup SI durations, namely, 4 days, 8 days, and 12 days.
Our experimental results (omitted for brevity)
show that the estimation accuracy is somewhat insensitive to the duration of the SI period.
Hence, for the experiments conducted for this paper, we chose 8 days as the SI period.

%=====

\section{An Additional Experiment for M2}

\subsection{Data Generation}
\label{sec:supp-datagen}
For this experiment, we use a different and slightly more general contact tracing graph to simulate the spread of infection. Recall that the adjacency matrix of the contact graph has a block diagonal structure with sizes of cliques coming from the distribution of family sizes in India \cite[pg. 18]{UNDoc}. However, in this case, we allow two consecutive (according to the order in which cliques appear along the diagonal of the contact matrix) non-trivial  cliques (i.e., cliques with more than one node) to have an overlap of one node with probability half. This assumption is reasonable since the concept of family encompasses more general groups such as people at the same workplace, students studying in the same classroom, etc. Furthermore, we remove $\alpha = 5\%$ of the edges from this block diagonal structure, thus converting the existing cliques into ``almost-cliques.'' This modified block diagonal structure is kept constant over time while the cross-clique contacts are updated every day. Except for the changes in the underlying contact tracing graph, the rest of the data generation method is the same as that described in Sec.~2 of the main paper.

\subsection{Inference}
We use the four algorithms (including \textsc{Comp}) for multiplicative noise described in Sec.~3 of the main paper. However, instead of using maximal cliques as groups in \textsc{Comp-sqrt-oglasso}, we use the decomposition of the contact tracing graph into overlapping 3-clique communities \cite{Palla2005}. An algorithm for detecting $k$-clique communities can be found in Sec.~1 of \cite[Supplementary Notes]{Palla2005}. The first step of this algorithm involves finding the maximal cliques in the contact graph, for which we use the Bron-Kerbosch algorithm \cite{Bron1973}. In the next step, we detect 3-clique communities and label each of those as groups. Further, we also label as groups the maximal cliques 
that are not part of any of these communities, in order to ensure that every contact is taken into account. The advantage of using 3-clique communities over just maximal cliques is that the former is able to capture ``almost cliques," i.e., cliques with a small fraction of absent pairwise contacts.

\subsection{Numerical Results} 
\begin{figure*}[!t]
\begin{minipage}[b]{.24\linewidth}
  \centering
  \centerline{\includegraphics[width=4.2cm]{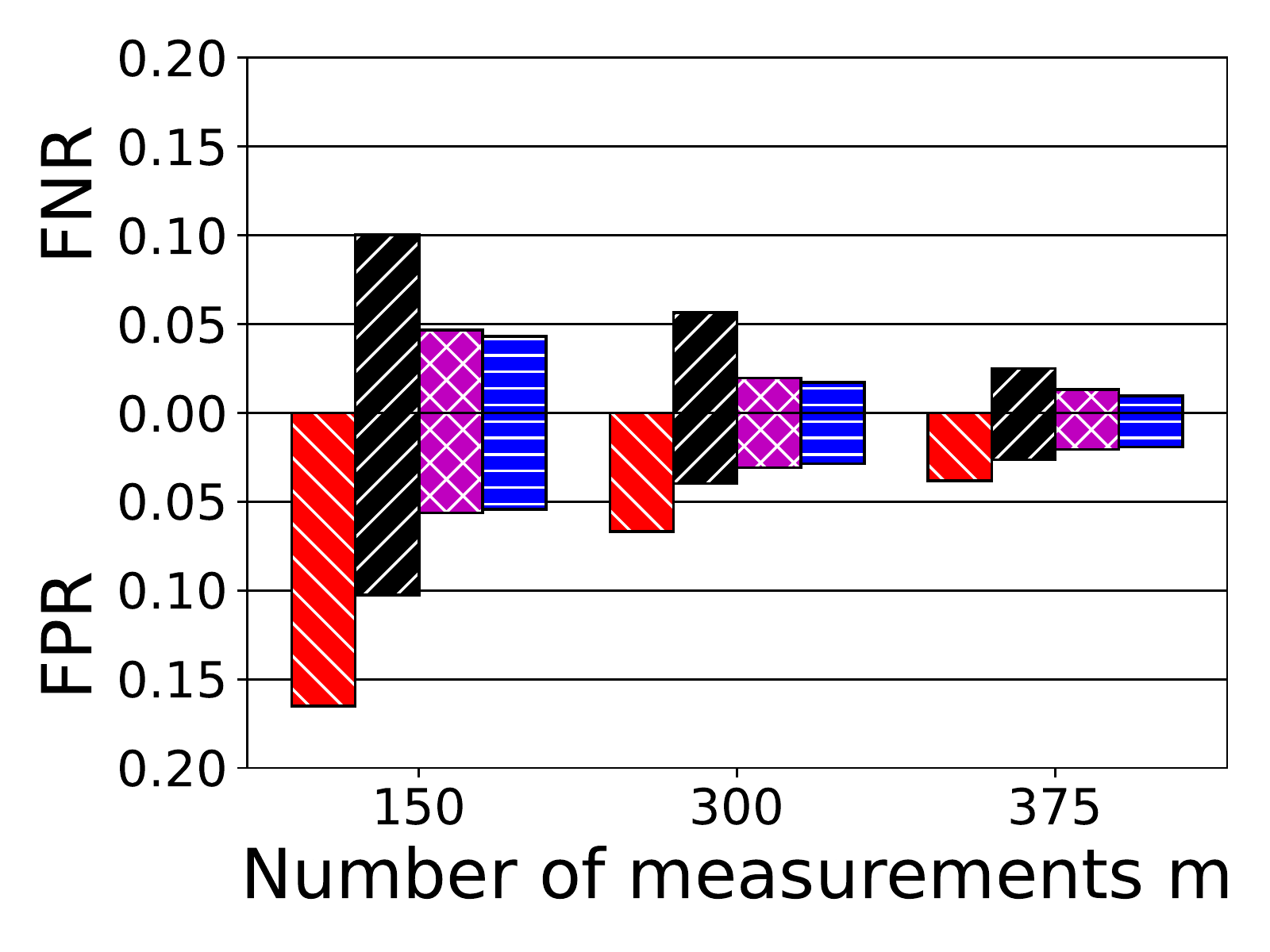}}
\end{minipage}
\hfill
\begin{minipage}[b]{0.24\linewidth}
  \centering
  \centerline{\includegraphics[width=4.2cm]{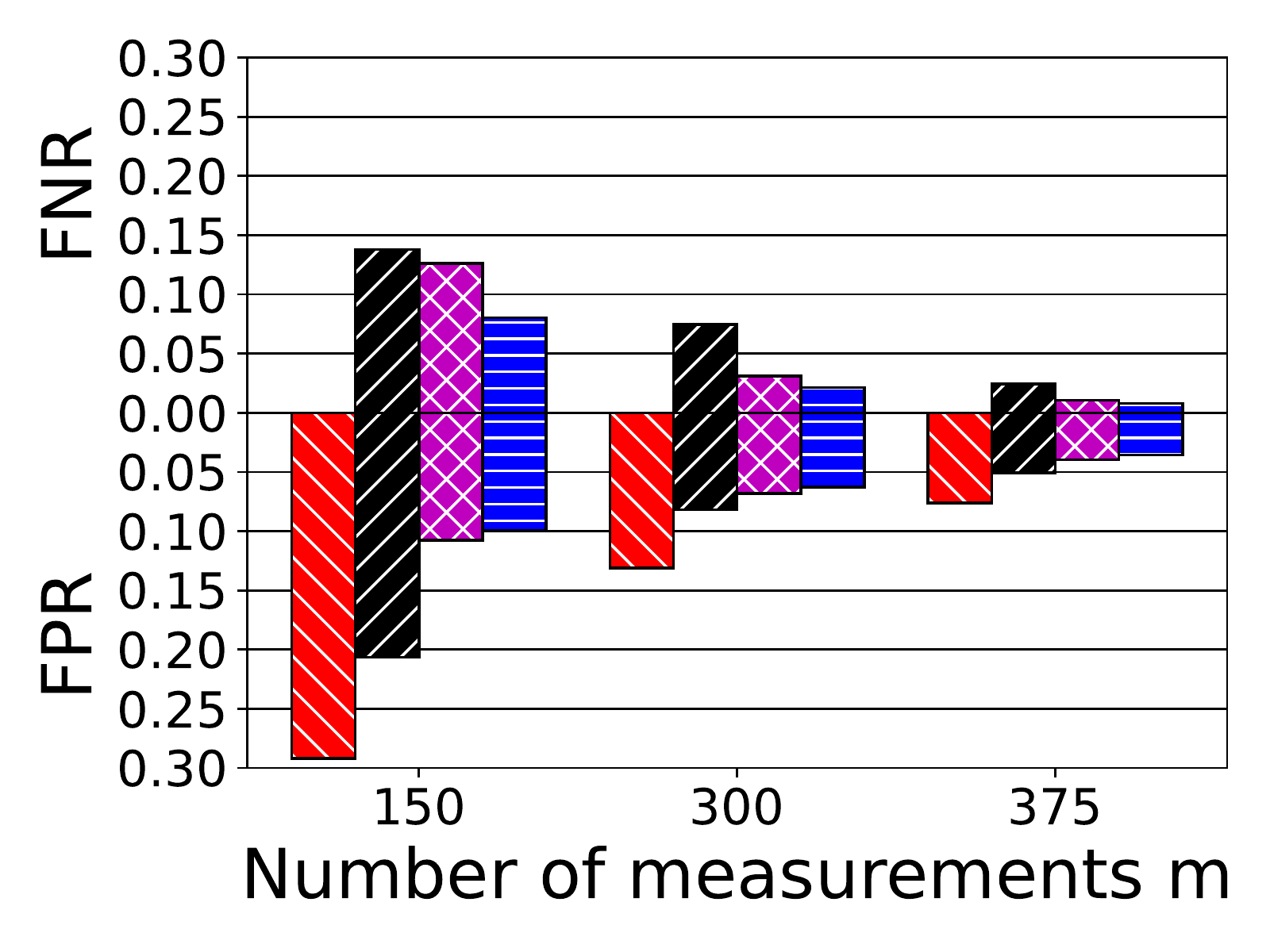}}
\end{minipage}
\hfill
\begin{minipage}[b]{.24\linewidth}
  \centering
  \centerline{\includegraphics[width=4.2cm]{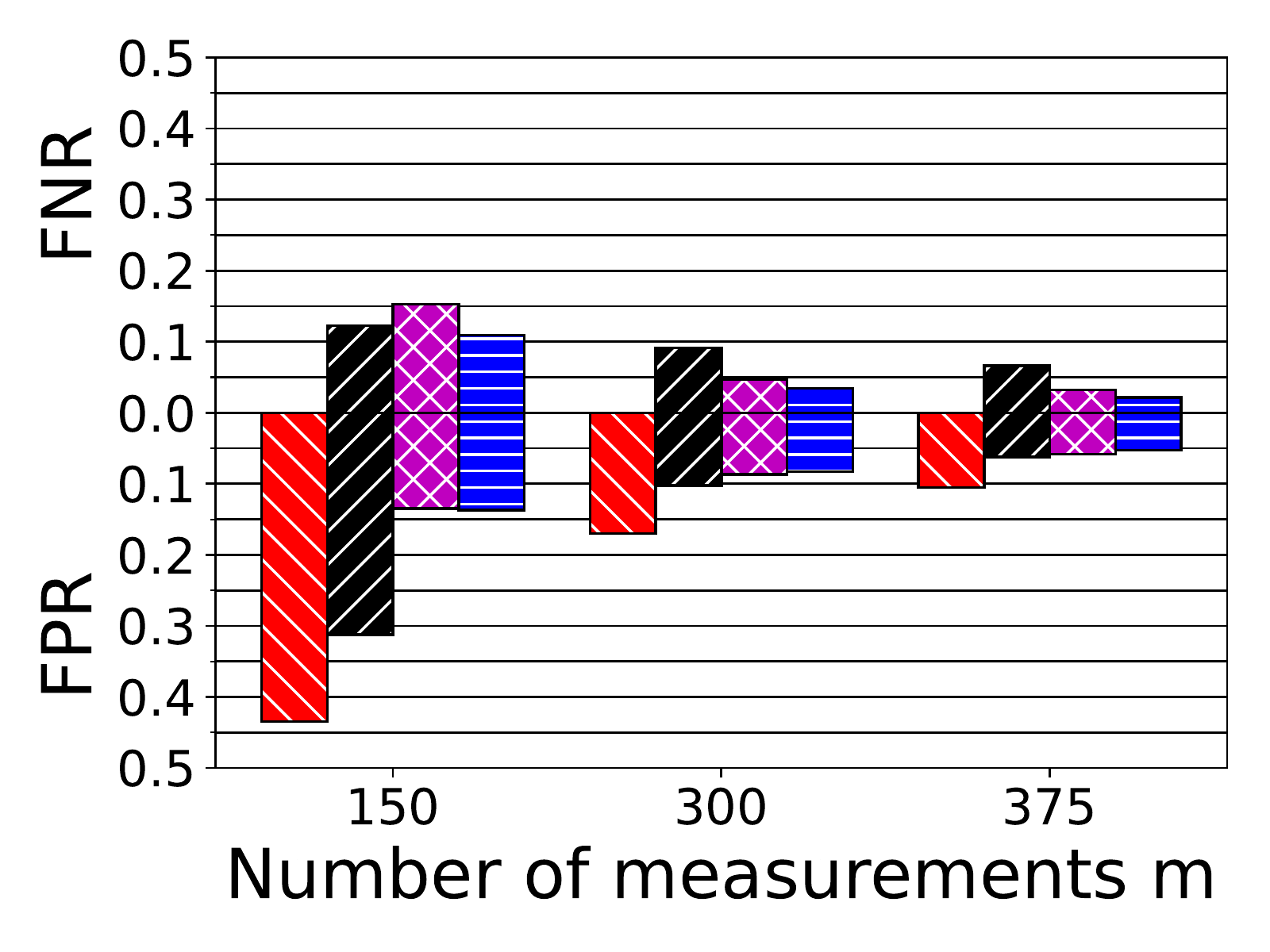}}
\end{minipage}
\hfill
\begin{minipage}[b]{0.24\linewidth}
  \centering
  \centerline{\includegraphics[width=4.2cm]{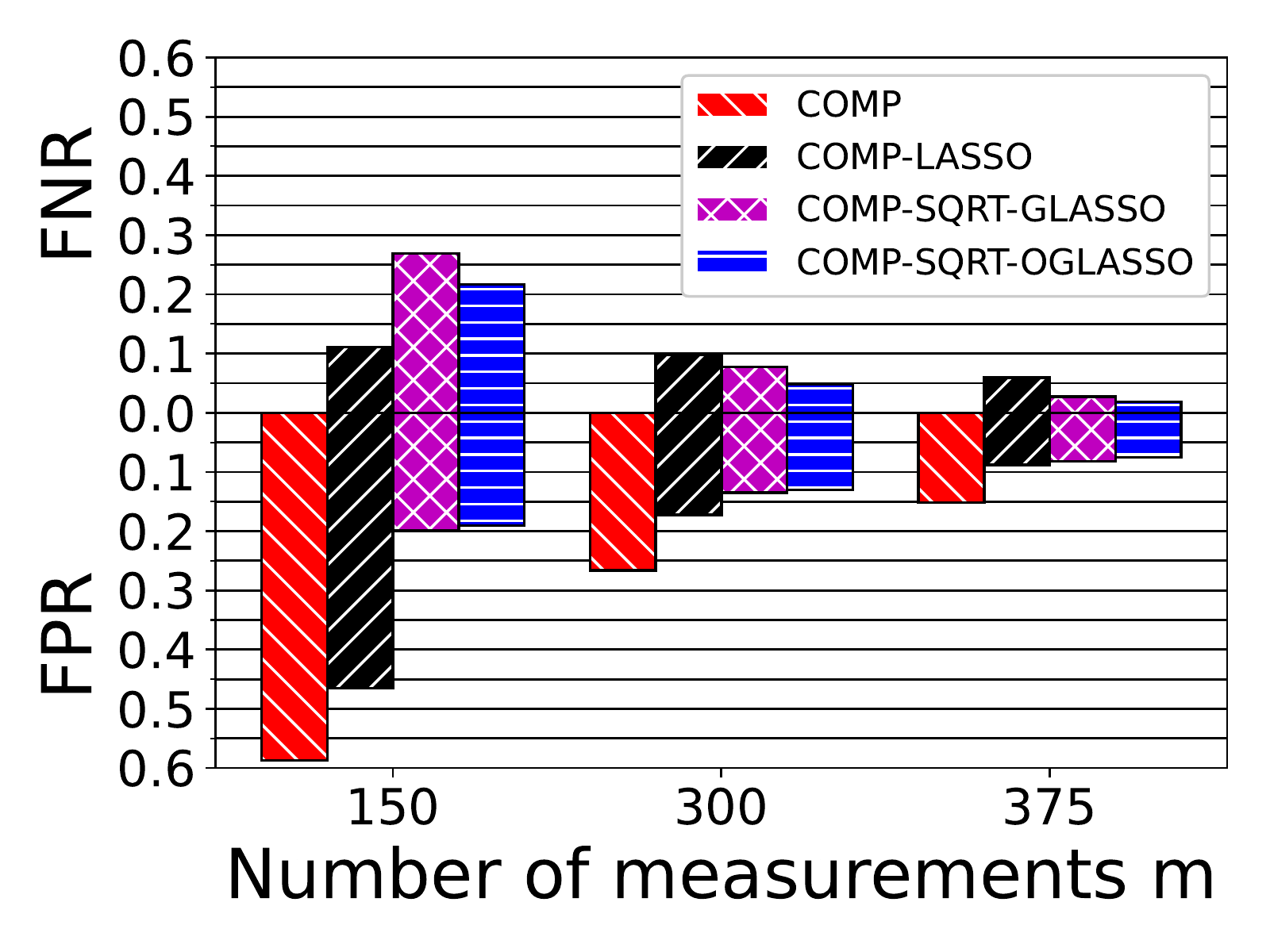}}
\end{minipage}

\caption{Figure showing mean FNR and FPR values for the contact graph from Section \ref{sec:supp-datagen}, for mean sparsity levels of 3.20\%, 4.84\%, 6.25\%, 8.66\% (from left to right).}
\label{fig:res-gen-m2}

\end{figure*}

\begin{figure*}[!t]
\begin{minipage}[b]{.24\linewidth}
  \centering
  \centerline{\includegraphics[width=4.35cm]{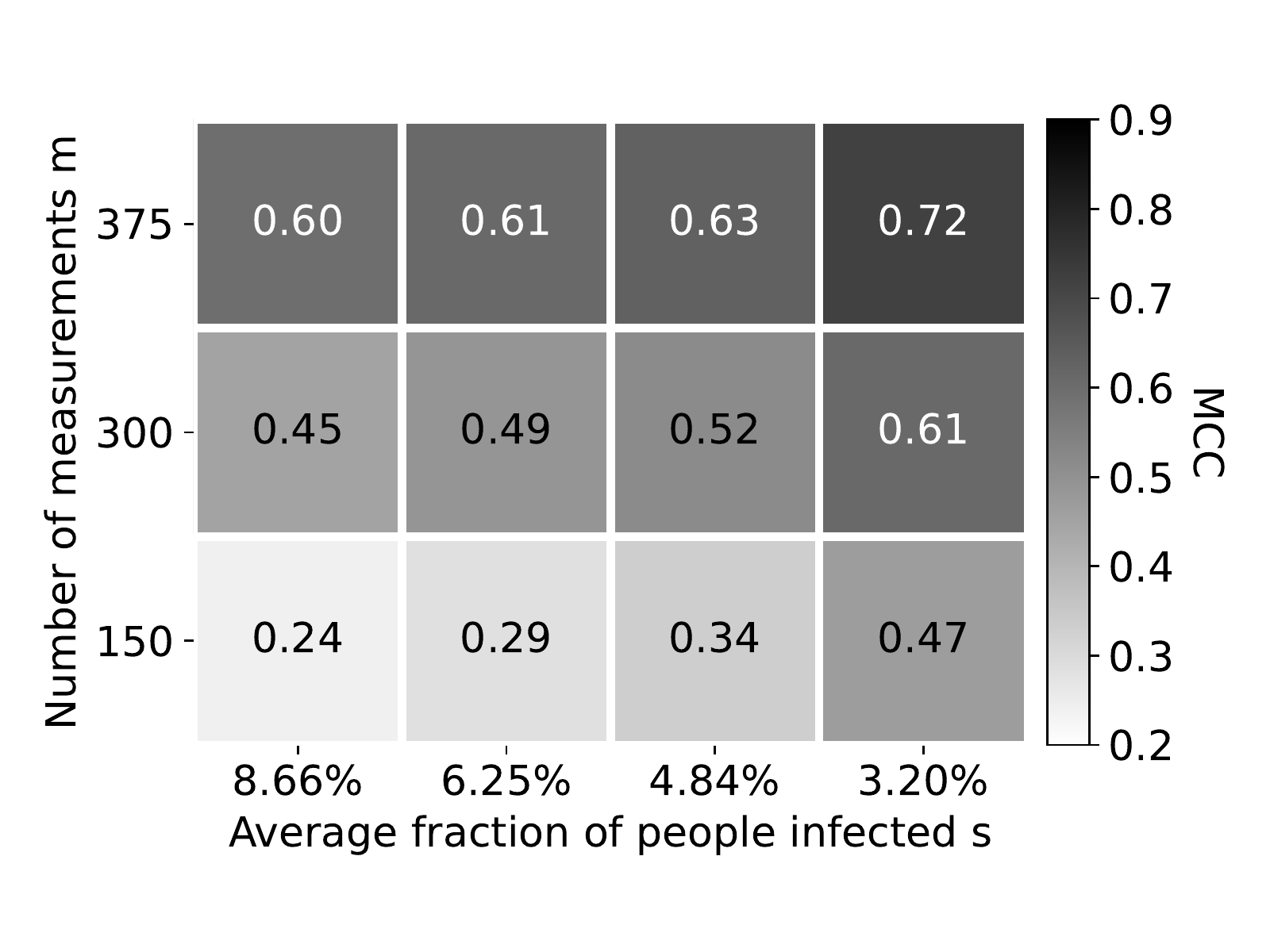}}
\end{minipage}
\hfill
\begin{minipage}[b]{0.24\linewidth}
  \centering
  \centerline{\includegraphics[width=4.35cm]{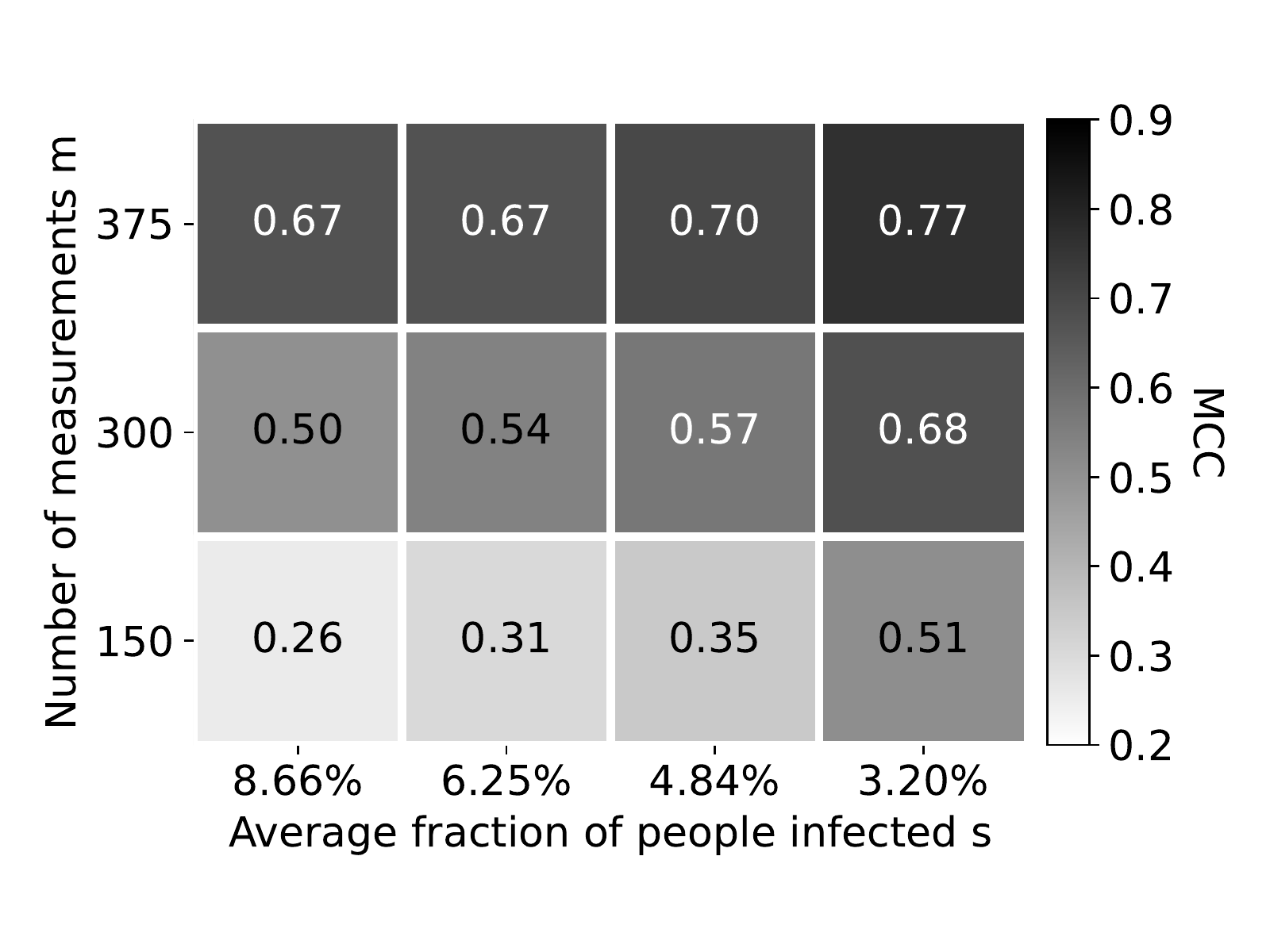}}
\end{minipage}
\hfill
\begin{minipage}[b]{.24\linewidth}
  \centering
  \centerline{\includegraphics[width=4.35cm]{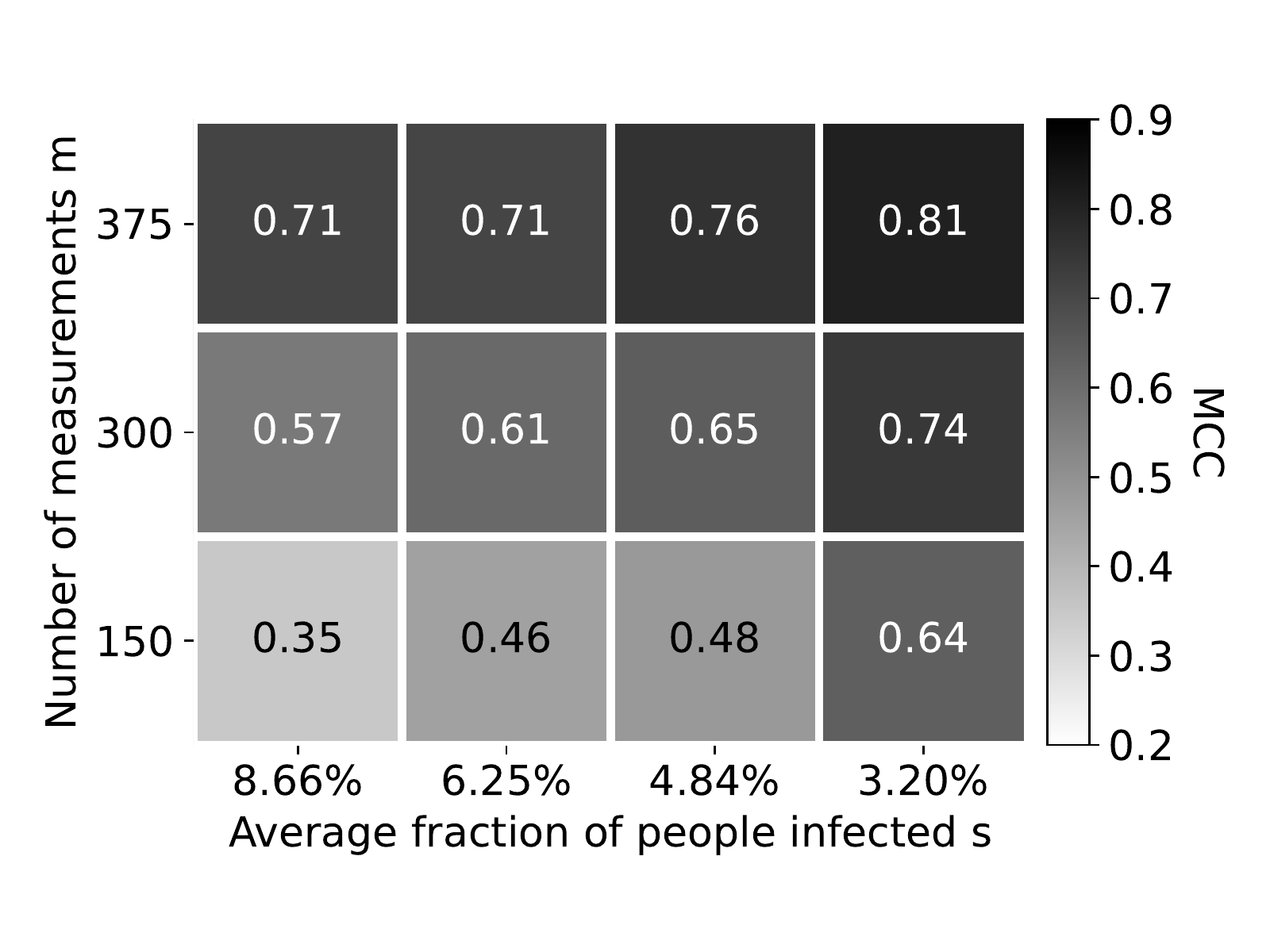}}
\end{minipage}
\hfill
\begin{minipage}[b]{0.24\linewidth}
  \centering
  \centerline{\includegraphics[width=4.35cm]{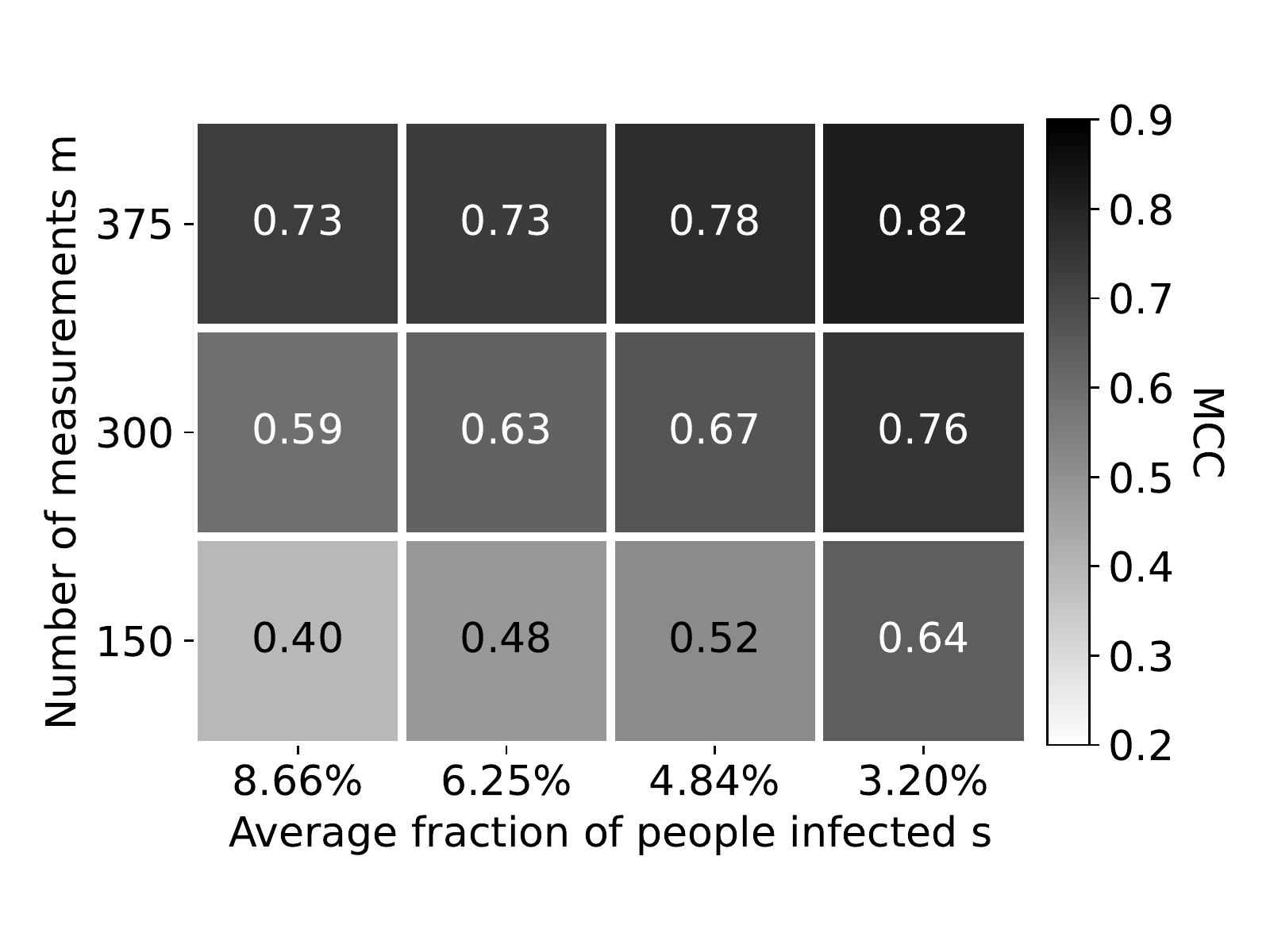}}
\end{minipage}

\caption{Figure showing mean MCC values obtained using \comp{}, \complasso{}, \compsqrtglasso{}, \compsqrtoglasso{} (from left to right).}
\label{fig:res-gen-mcc-m2}

\end{figure*}

\begin{figure*}[!t]
\begin{minipage}[b]{.33\linewidth}
  \centering
  \centerline{\includegraphics[width=4.35cm]{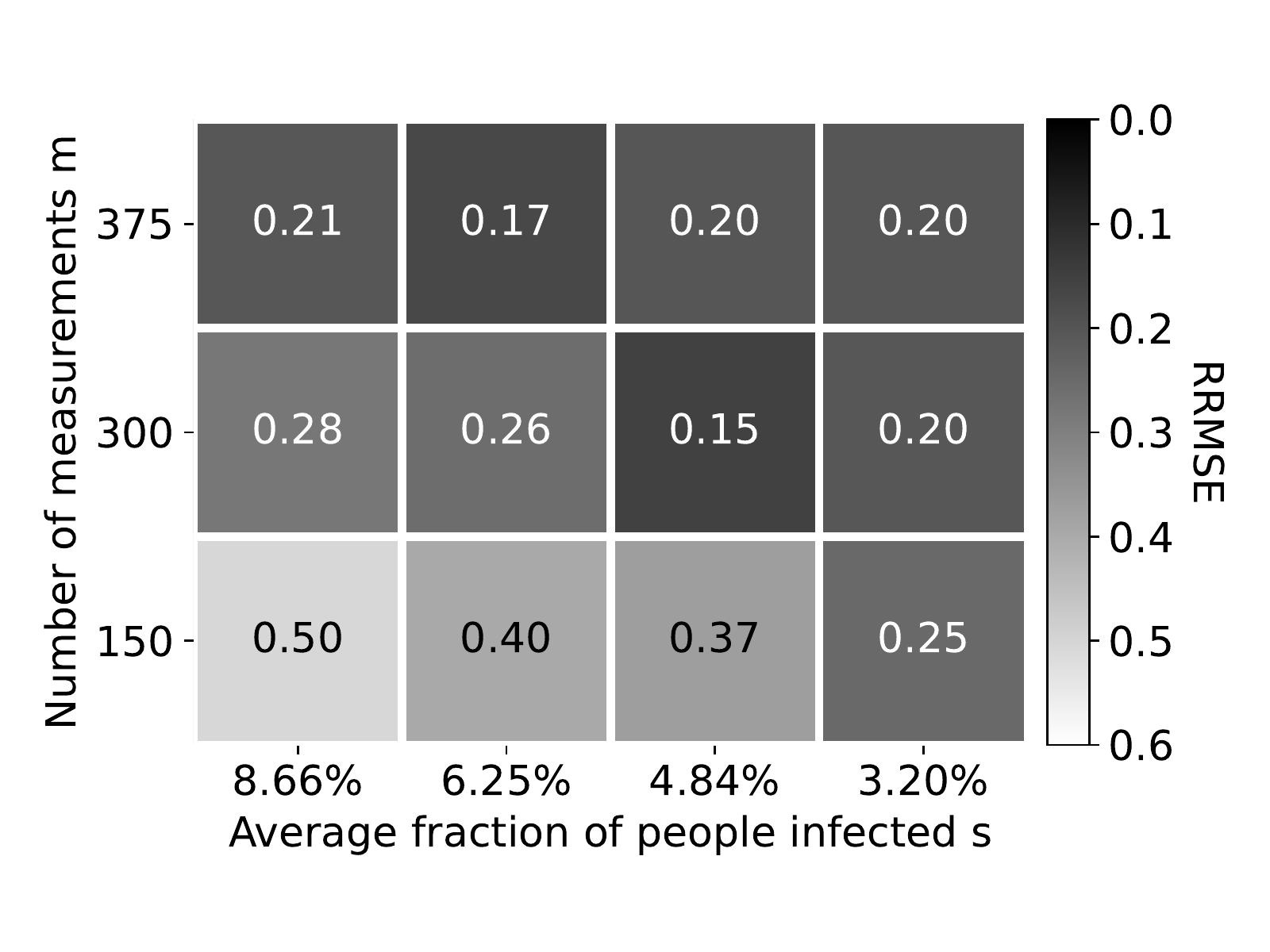}}
\end{minipage}
\hfill
\begin{minipage}[b]{0.33\linewidth}
  \centering
  \centerline{\includegraphics[width=4.35cm]{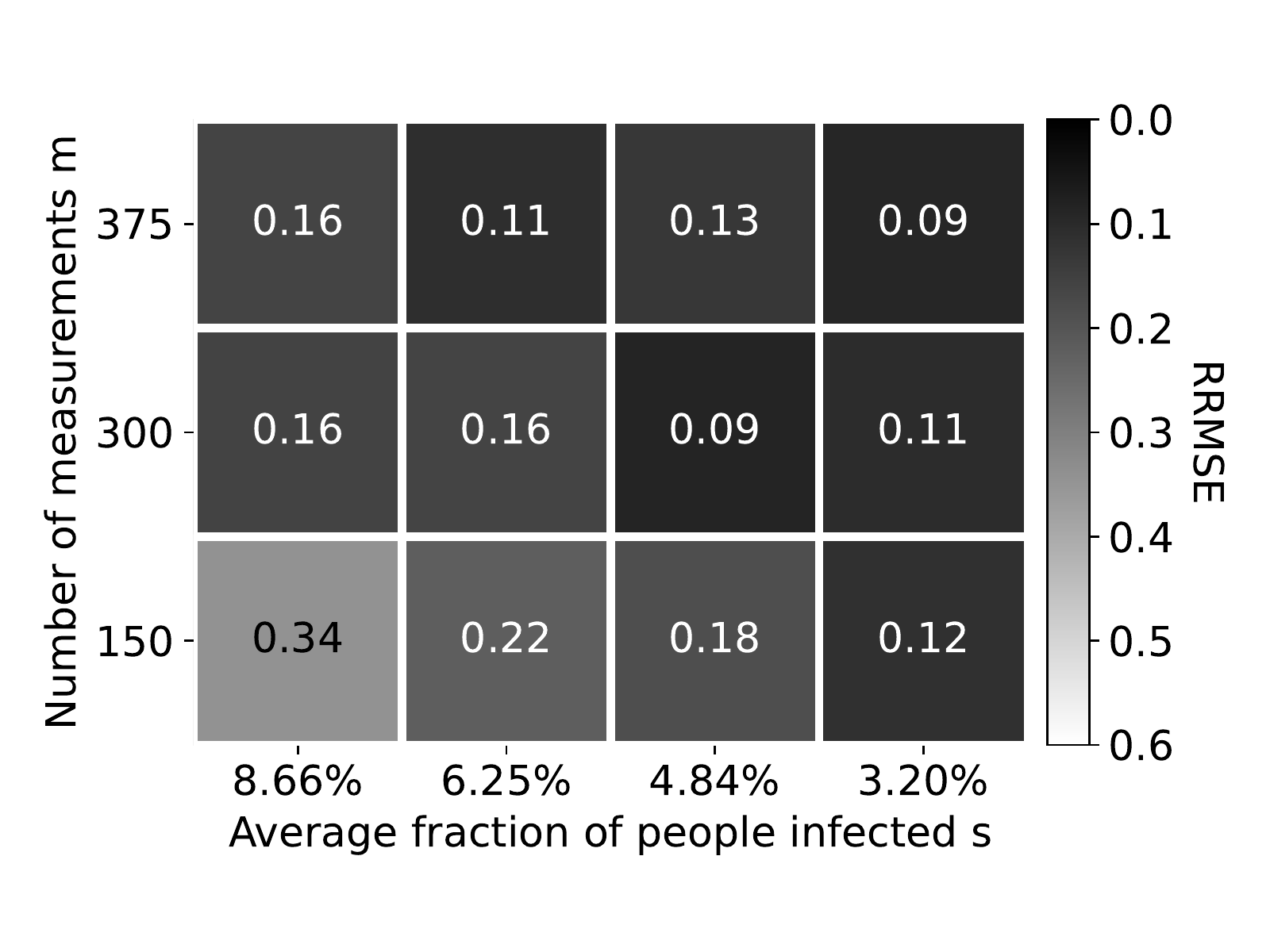}}
\end{minipage}
\hfill
\begin{minipage}[b]{.33\linewidth}
  \centering
  \centerline{\includegraphics[width=4.35cm]{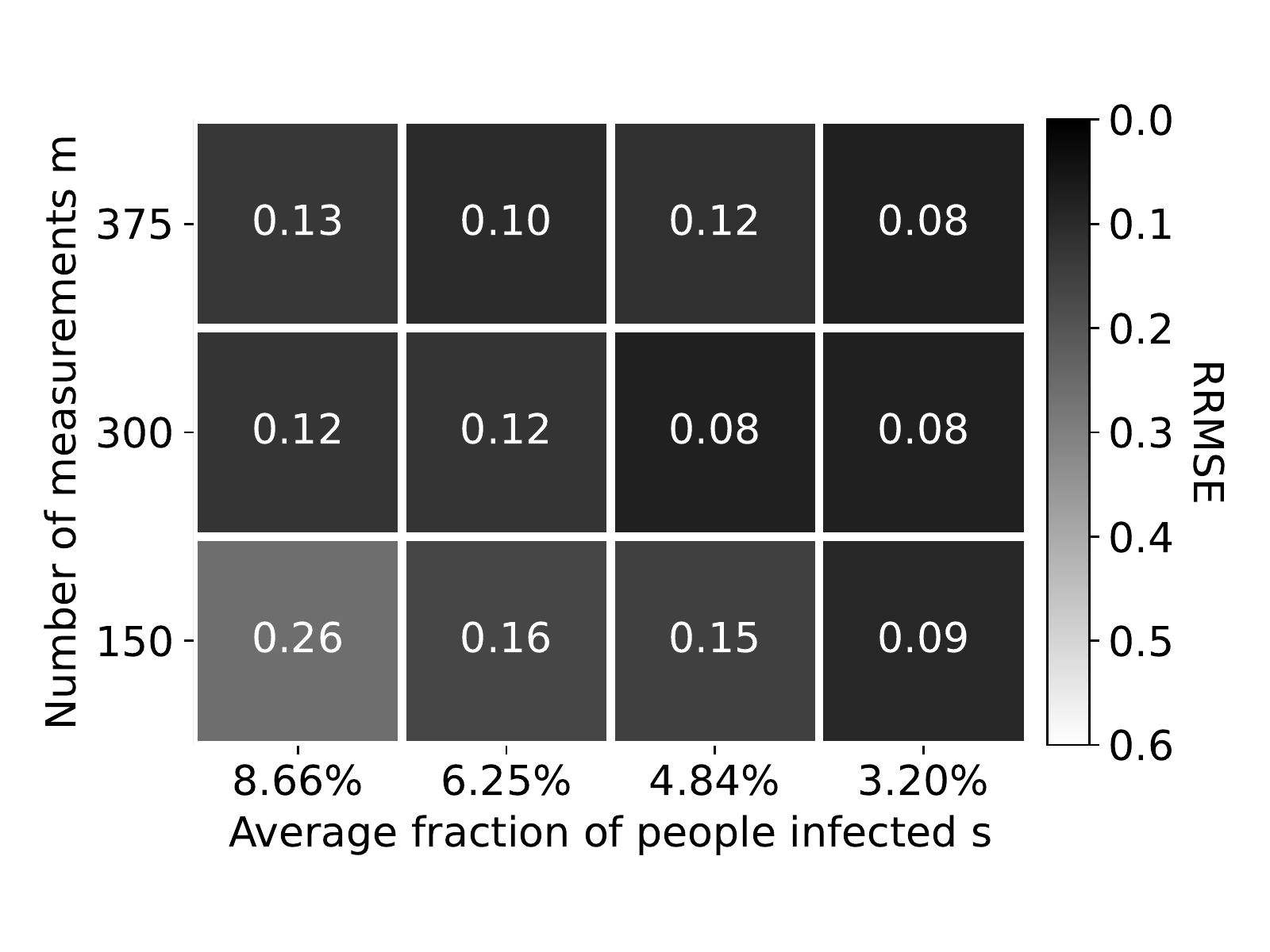}}
\end{minipage}

\caption{Figure showing mean RRMSE values obtained using \complasso{}, \compsqrtglasso{}, \compsqrtoglasso{} (from left to right).}
\label{fig:res-gen-rrmse-m2}

\end{figure*}

We present the results in a format similar to that in Sec.~4 of the main paper, but for the contact graph described in Sec.~\ref{sec:supp-datagen}. Fig.~\ref{fig:res-gen-m2} shows the mean values (across 50 signals) of the
false negative rate (FNR) and false positive rate (FPR) obtained for four different sparsity levels. The sparsity levels were obtained by varying the amount of cross-clique contacts. We remark that the length of each bar in Fig.~\ref{fig:res-gen-m2}, $\text{FNR} + \text{FPR}$, is equal to $1 - \text{Youden's Index}$.
Further, we plot heat maps (Fig.~\ref{fig:res-gen-mcc-m2}) to compare the performance of the four algorithms under consideration---the intensity of gray corresponds to the mean value (across 50 signals) of the Matthews Correlation Coefficient (MCC). The MCC is defined as
\begin{equation}
\text{MCC} = \dfrac{\text{TP} \times \text{TN} - \text{FP} \times \text{FN}}{\sqrt{(\text{TP}\!+\!\text{FP})(\text{TP}\!+\!\text{FN})(\text{TN}\!+\!\text{FP})(\text{TN}\!+\!\text{FN})}},
\end{equation}
and has been proposed as a comprehensive metric to evaluate the performance of binary classification algorithms \cite{Chicco2020}. Its values range from $-1$ to $+1$, where a value closer toward $+1$ is desirable. The RRMSE values can be seen from the heat maps in Fig.~\ref{fig:res-gen-rrmse-m2} (we do not provide a heat map for \textsc{Comp} since it does not estimate viral loads).

\section{Additional Results for M2}

\begin{figure*}[!t]
\begin{minipage}[b]{.24\linewidth}
  \centering
  \centerline{\includegraphics[width=4.35cm]{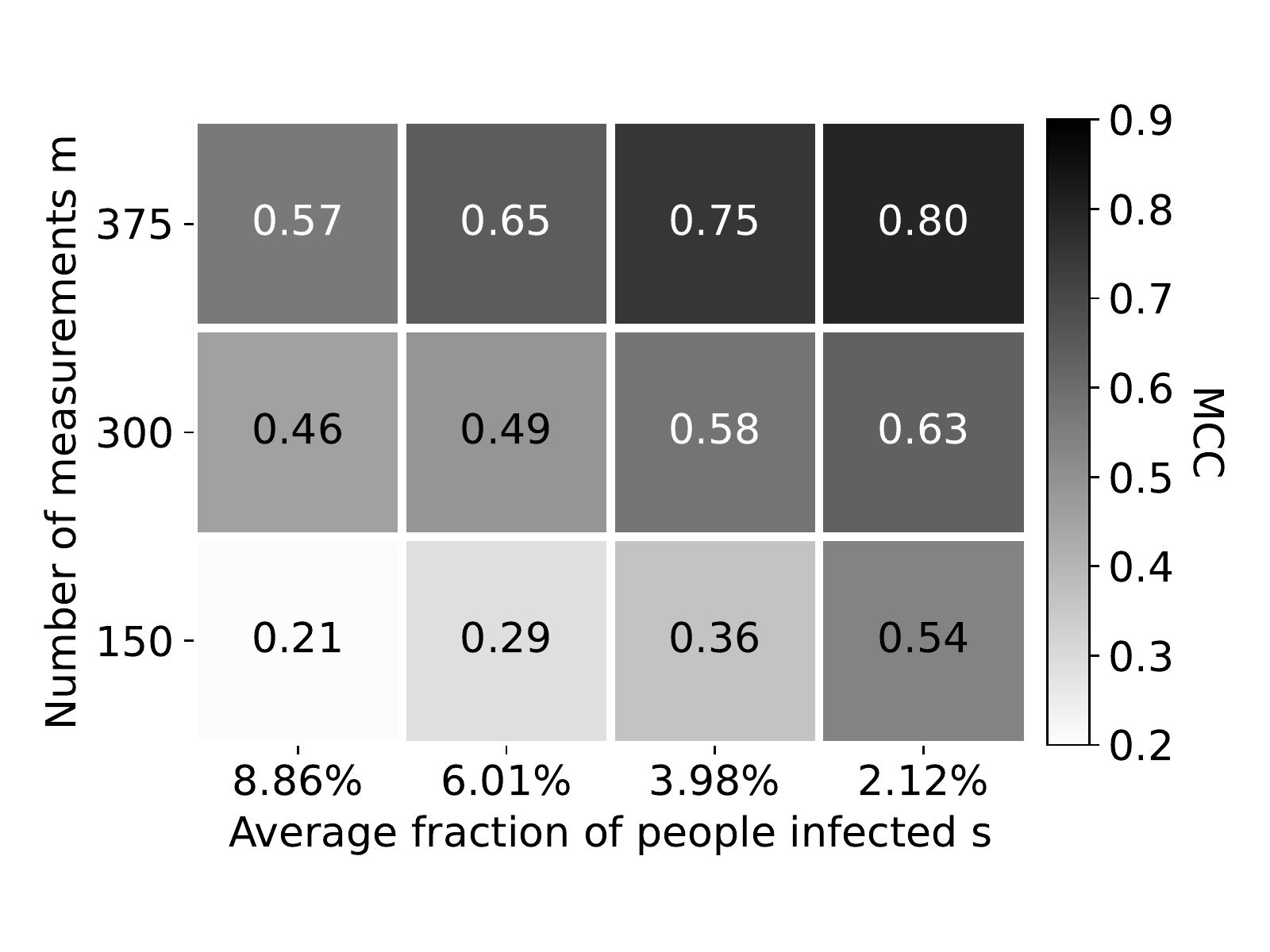}}
\end{minipage}
\hfill
\begin{minipage}[b]{0.24\linewidth}
  \centering
  \centerline{\includegraphics[width=4.35cm]{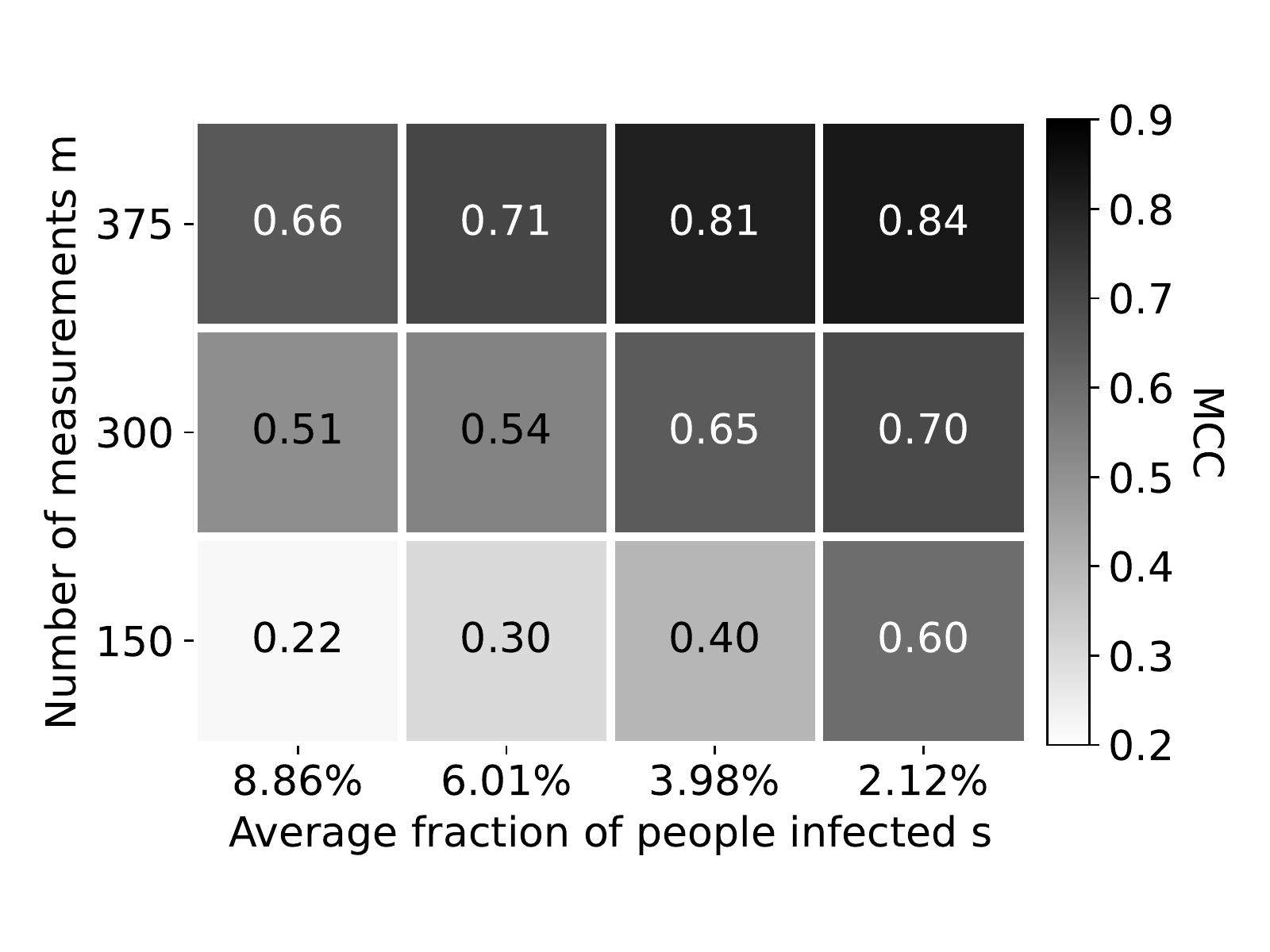}}
\end{minipage}
\hfill
\begin{minipage}[b]{.24\linewidth}
  \centering
  \centerline{\includegraphics[width=4.35cm]{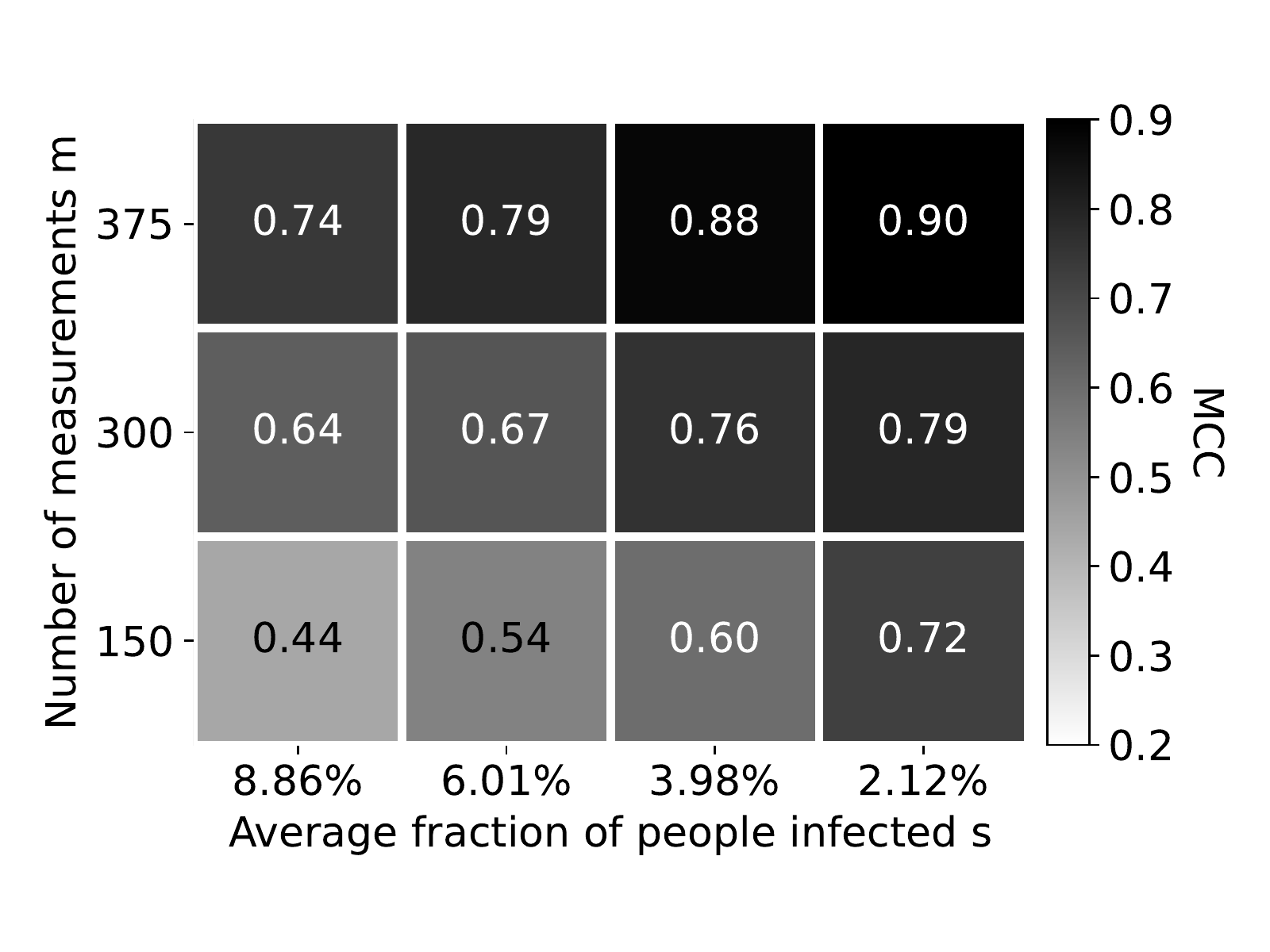}}
\end{minipage}
\hfill
\begin{minipage}[b]{0.24\linewidth}
  \centering
  \centerline{\includegraphics[width=4.35cm]{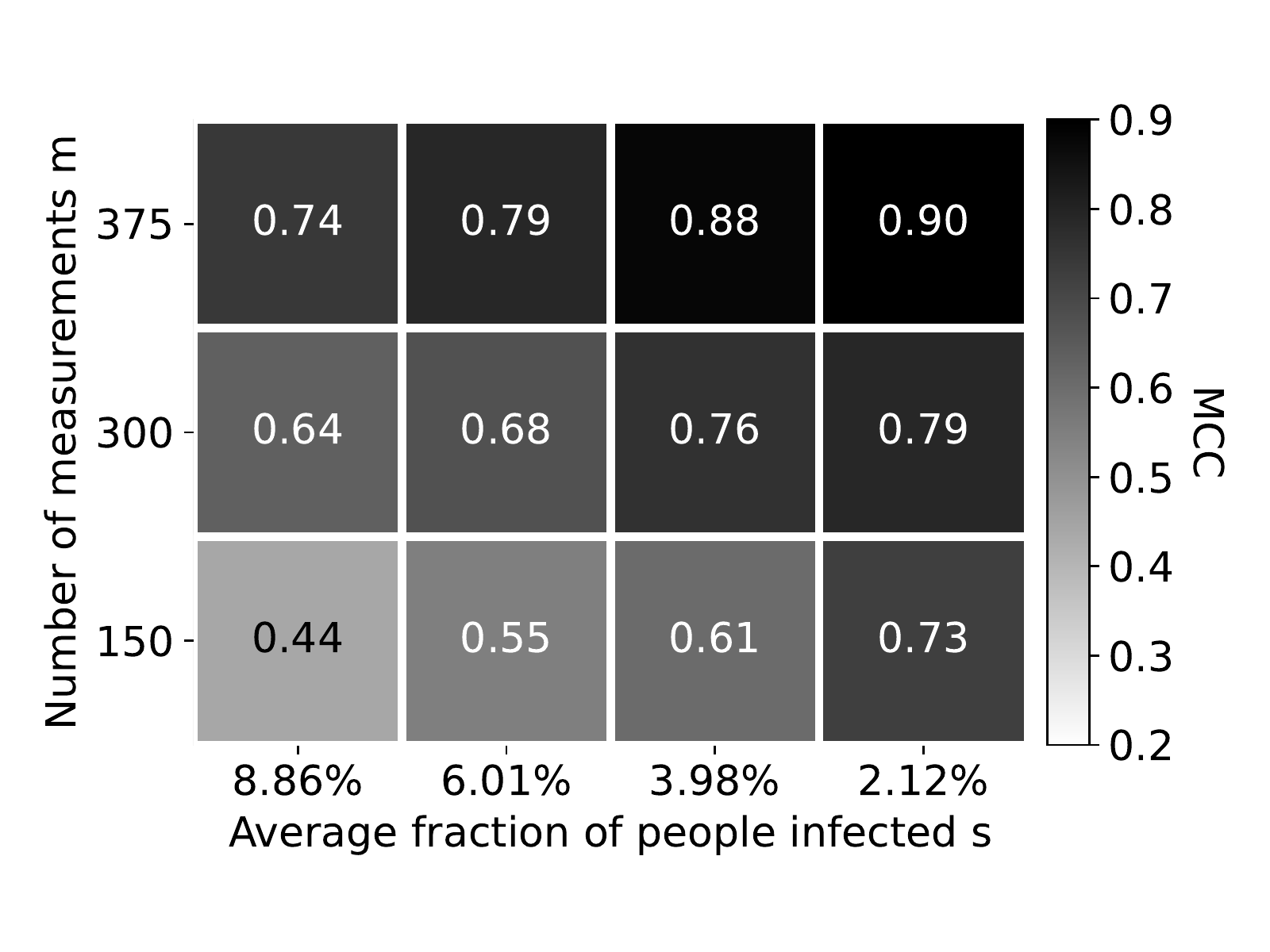}}
\end{minipage}

\caption{Figure showing mean MCC values obtained using \comp{}, \complasso{}, \compsqrtglasso{}, \compsqrtoglasso{} (from left to right).}
\label{fig:res-mcc-m2}

\end{figure*}

\begin{figure*}[!t]
\begin{minipage}[b]{.33\linewidth}
  \centering
  \centerline{\includegraphics[width=4.35cm]{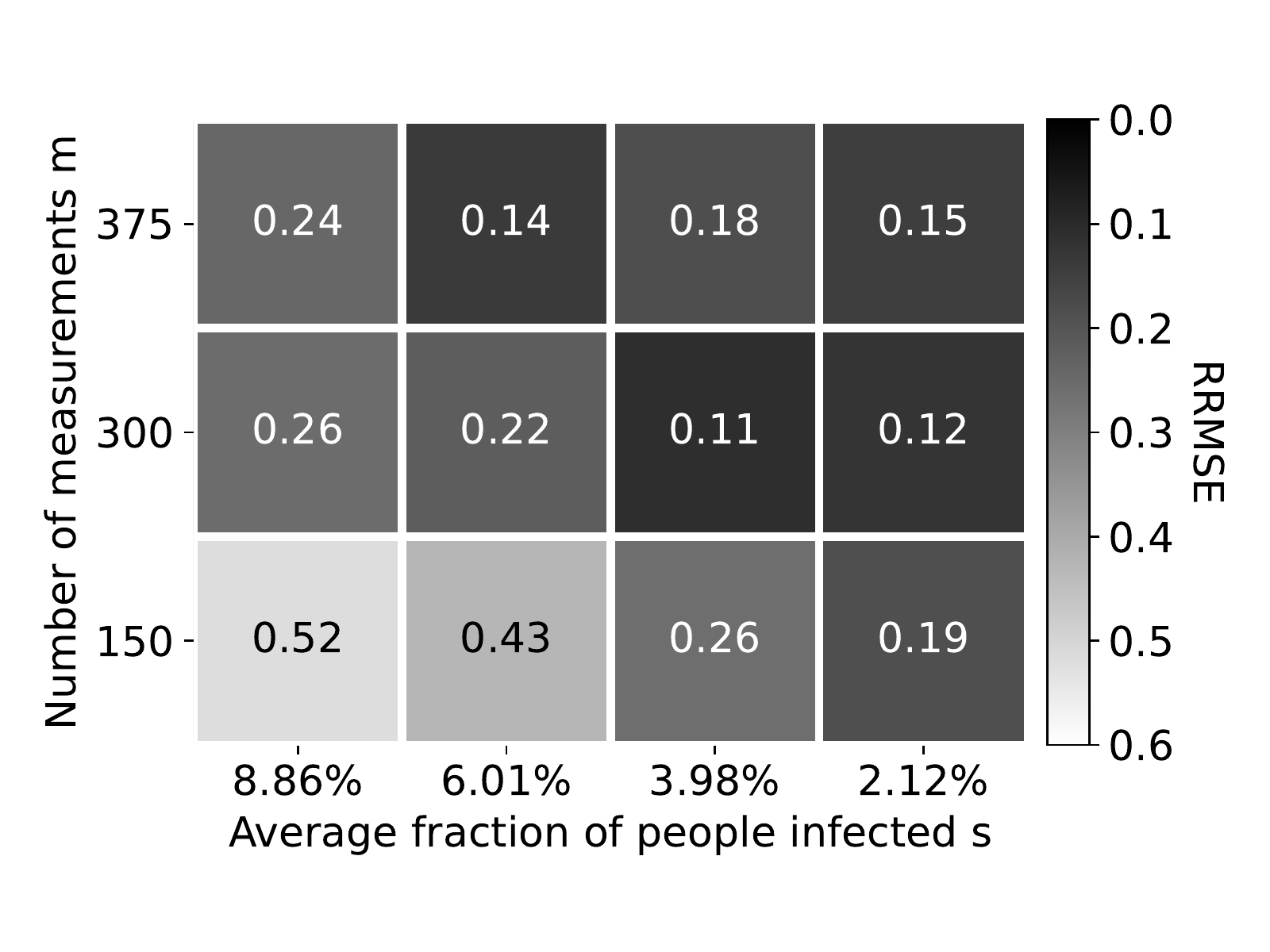}}
\end{minipage}
\hfill
\begin{minipage}[b]{0.33\linewidth}
  \centering
  \centerline{\includegraphics[width=4.35cm]{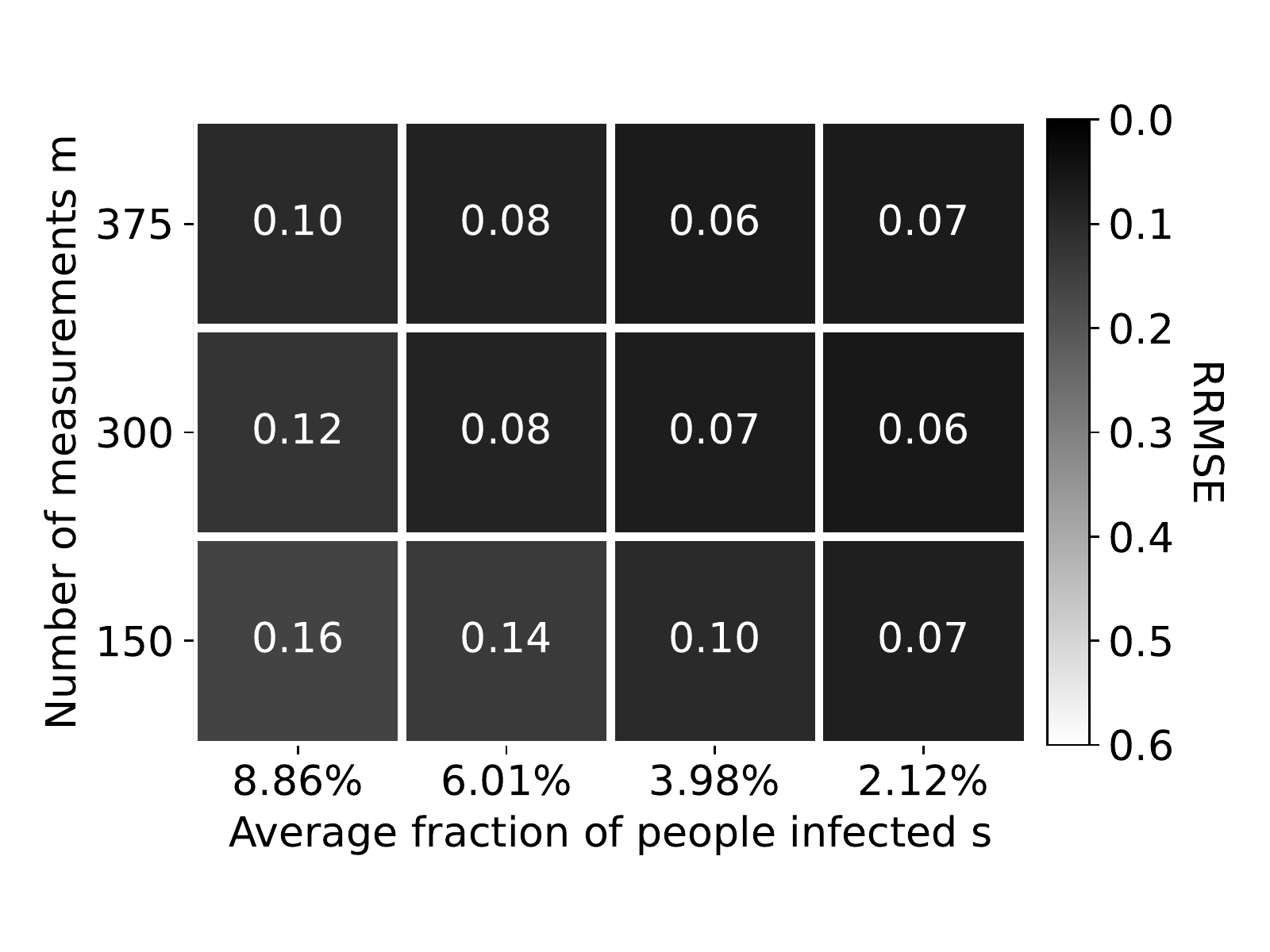}}
\end{minipage}
\hfill
\begin{minipage}[b]{.33\linewidth}
  \centering
  \centerline{\includegraphics[width=4.35cm]{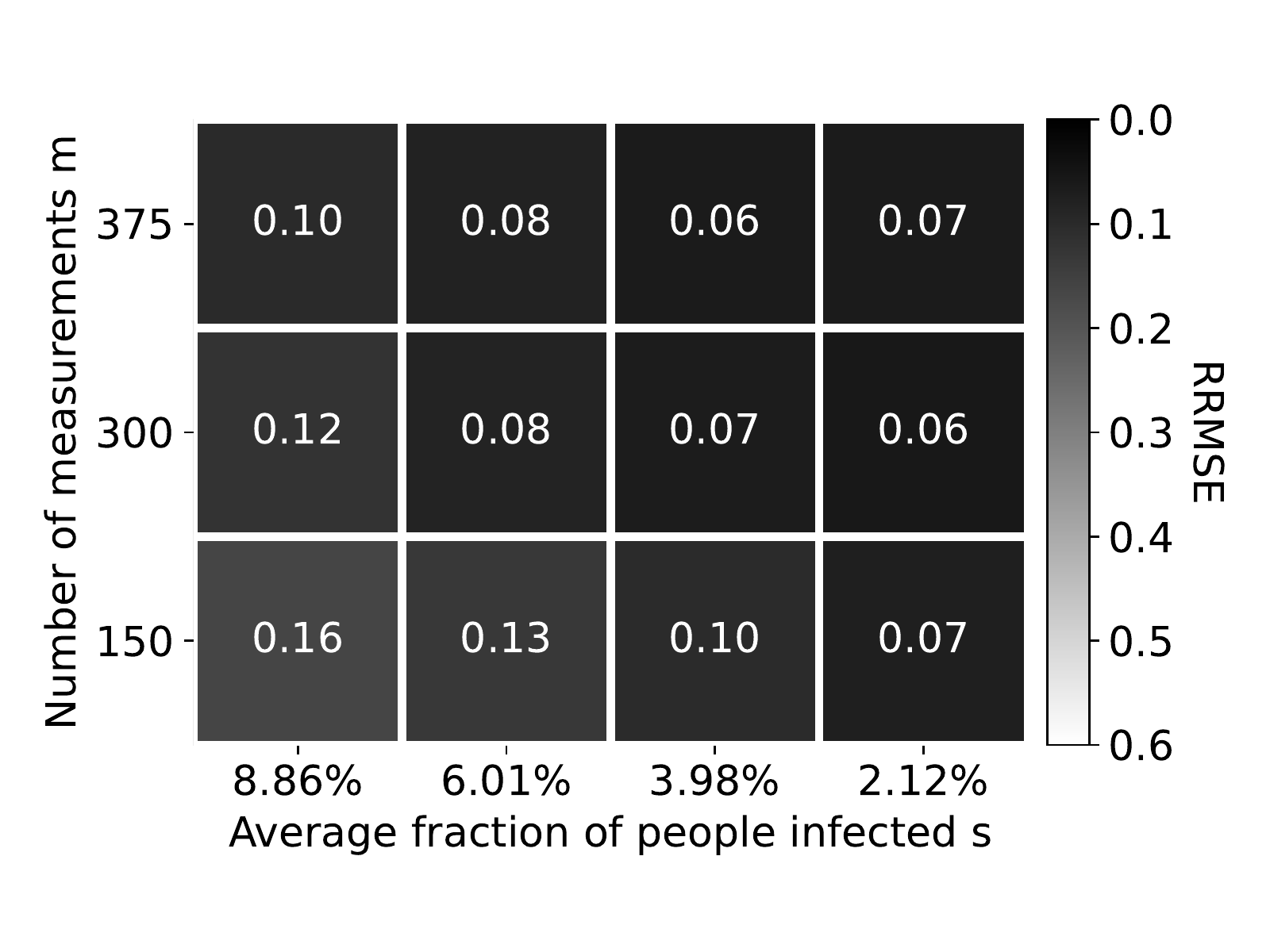}}
\end{minipage}

\caption{Figure showing mean RRMSE values obtained using \complasso{}, \compsqrtglasso{}, \compsqrtoglasso{} (from left to right).}
\label{fig:res-rrmse-m2}

\end{figure*}

For model \textbf{M2}, we present a comparison of the four algorithms for the experiment described in Sec. 4 of the main paper. Fig.~\ref{fig:res-mcc-m2} and Fig.~\ref{fig:res-rrmse-m2} show a comparison of the performance of the algorithms under consideration in terms of mean MCC and mean RRMSE values, respectively. Further, we remark that the true viral loads of the false negatives yielded by \complasso{} variants are concentrated toward lower values. For instance, only about 29\% of the false negatives given by \compsqrtoglasso{} had viral load values greater than $2^{12} = 4096$.

\section{Sensing Matrix Design}
As mentioned in the main paper, we use Kirkman triple matrices as sensing matrices for performing pooling. A Kirkman triple (binary) matrix $\boldsymbol{A}$ can be partitioned into $3n/m$ sub-matrices of dimensions $m \times m/3$, each of which contains exactly one nonzero entry in each row and three nonzero entries in each column. Further, the dot product of any two columns of the matrix $\boldsymbol{A}$ should not exceed 1. For a given value of $n$, $m (< n)$ must satisfy the following conditions:
\begin{enumerate}
    \item $m$ must be of the form $3n_1$, where $n_1$ divides $n$, since the number of sub-matrices and the number of columns in each sub-matrix must be integers.
    \item $\binom{3}{2} \cdot n \leq m(m-1)/2$ since a triple contains $\binom{3}{2}$ pairs and a pair must belong to at most one triple.
\end{enumerate}
For $n = 1000$, the only values of $m$ which satisfy the above constraints are $120$, $150$, $300$, $375$, $600$, and $750$. We construct Kirkman triple matrices with $m = 150, 300, 375$ and use them in our experiments. 
The matrices are constructed based on a few simple rules:
\begin{enumerate}
    \item The indices of ones in each column form an arithmetic progression (AP).
    \item The matrix has a block structure and the common difference of the AP remains constant throughout each block. Furthermore, the sum of all columns in a block yields the vector consisting of all ones.
    \item The common difference values $\{d_B : \boldsymbol{B}\text{ is a block}\}$ are chosen such that the multi-set $\{r \cdot d_B: r \in \{n: n \in \mathbb{N}, n < 3\}, \boldsymbol{B} \text{ is a block}\}$ has no duplicate values.
\end{enumerate}

\begin{figure*}[!t]
  \centering
  \centerline{\includegraphics[width=16cm]{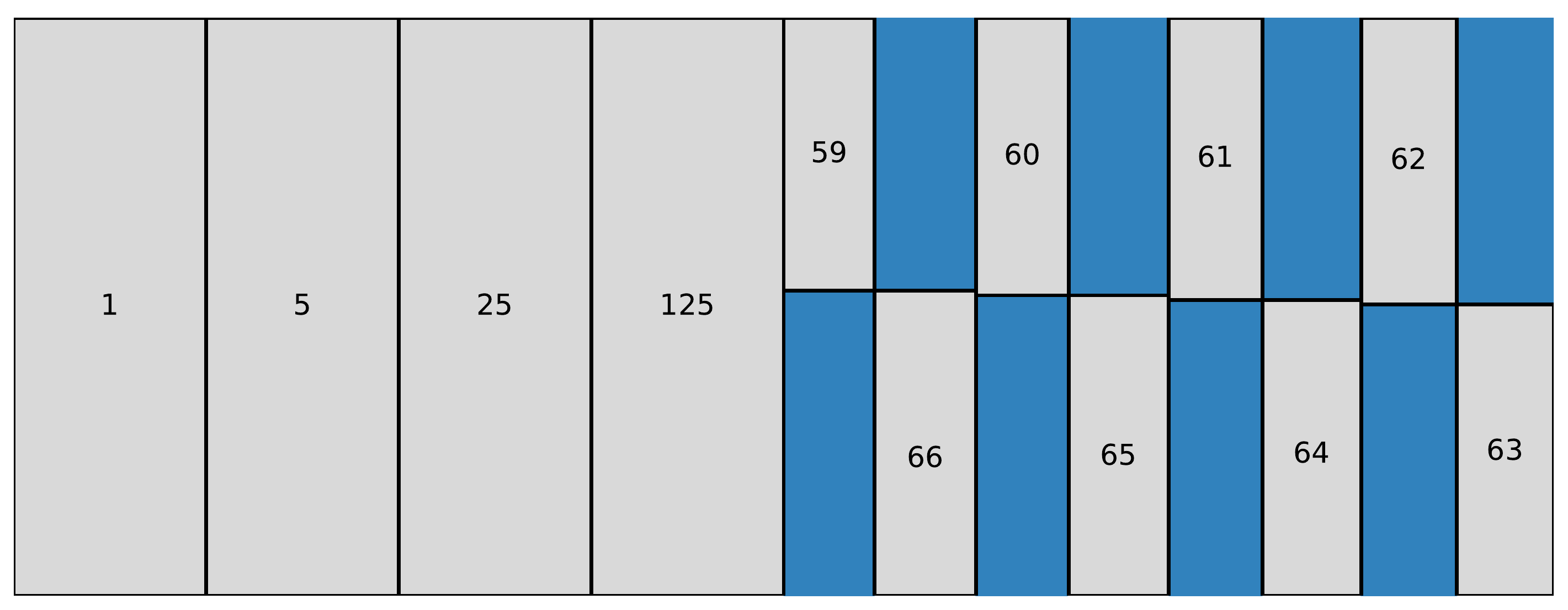}}

\caption{A $375 \times 1000$ Kirkman triple matrix obtained using our approach. The number written within each block is equal to the corresponding common difference value and the blocks without number markings are zero matrices. The blocks having less than $m = 375$ rows have dimensions equal to $3d_B \times d_B$.}
\label{fig:375-1000-kirkman}
\end{figure*}
Fig.~\ref{fig:375-1000-kirkman} shows the structure of a $375 \times 1000$ Kirkman triple matrix obtained using the above approach. Let $\boldsymbol{B}$ be any block and let $d_B$ denote the common difference of the AP for block $\boldsymbol{B}$ as indicated in the figure. Then, the $i$th column of $\boldsymbol{B}$ is given by
\begin{equation}
    \boldsymbol{B}_i = \sum_{j=1}^{3-1} \boldsymbol{e}_{\beta_i + j d_B},\quad \beta_i = \text{mod}(i,d_B) + 3d_B[i/d_B],
\end{equation}
where $[\cdot]$ denoted the greatest integer function and $\boldsymbol{e}_j$ denotes the $j$th standard basis vector. Clearly, any block $\boldsymbol{B}$ must have dimensions $3 n_B \times n_B$, where $d_B$ divides $n_B$.

As $m$ decreases, it becomes harder to design matrices satisfying all three rules specified earlier. However, it is possible to relax the third rule in such cases and still obtain a matrix satisfying the required constraints. For example, our $150 \times 1000$ Kirkman triple matrix does not obey the third rule. We further remark that one may design balanced matrices with a different number of (say $k$) ones in each column such that the dot product of every pair of columns is bounded by 1, using the above approach. Such matrices would arise from the Steiner systems $S(2,k,m)$ [just as Kirkman matrices arise from $S(2,3,m)$]. For example, it is straightforward to design a $400 \times 1000$ matrix with $k = 4$ using our approach.

\end{document}